\documentclass[twocolumn]{aastex631}

\usepackage{amsmath,amsfonts,amsthm,bm}
\usepackage{csquotes}

\begin{document}

\title{The All-Sky Impact of the LMC on the Milky Way Circumgalactic Medium}

\correspondingauthor{Christopher Carr}
\email{cc4504@columbia.edu}

\author[0000-0002-0786-7307]{Christopher Carr}
\affiliation{Department of Astronomy, Columbia University, 550 West 120th Street, New York, NY, 10027, USA} 

\author[0000-0003-2630-9228]{Greg L. Bryan}
\affiliation{Department of Astronomy, Columbia University, 550 West 120th Street, New York, NY, 10027, USA}

\author[0000-0001-7107-1744]{Nicolás Garavito-Camargo}
\affiliation{Center for Computational Astrophysics, Flatiron Institute, Simons Foundation, 162 Fifth Avenue, New York, NY 10010, USA}

\author[0000-0003-0715-2173]{Gurtina Besla}
\affil{Steward Observatory, University of Arizona, 933 North Cherry Avenue, Tucson, AZ 85721, USA}

\author[0000-0003-4075-7393]{David J. Setton}\thanks{Brinson Prize Fellow}
\affiliation{Department of Astrophysical Sciences, Princeton University, 4 Ivy Lane, Princeton, NJ 08544, USA}

\author[0000-0001-6244-6727]{Kathryn V. Johnston}
\affiliation{Department of Astronomy, Columbia University, 550 West 120th Street, New York, NY, 10027, USA}

\begin{abstract}
The first infall of the LMC into the Milky Way (MW) represents a large and recent disruption to the MW circumgalactic medium (CGM). In this work, we use idealized, hydrodynamical simulations of a MW-like CGM embedded in a live dark matter halo with an infalling LMC-like satellite initialized with its own CGM to understand how the encounter is shaping the global physical and kinematic properties of the MW CGM. First, we find that the LMC sources order-unity enhancements in MW CGM density, temperature, and pressure from a $\mathcal{M} \approx 2$ shock from the supersonic CGM-CGM collision, extending from the LMC to beyond $\sim R_{\rm 200, MW}$, enhancing column densities, X-ray brightness, the thermal Sunyaev-Zeldovich (tSZ) distortion, and potentially synchrotron emission from cosmic rays over large angular scales across the Southern Hemisphere. Second, the MW’s reflex motion relative to its outer halo produces a dipole in CGM radial velocities, with $v_{\rm R} \pm 30-50$ km/s at $R > 50$ kpc in the Northern/Southern hemispheres respectively, consistent with measurements in the stellar halo. Finally, ram pressure strips most of the LMC CGM gas by the present day, leaving $\sim 10^{8-9} M_{\odot}$ of warm, ionized gas along the past orbit of the LMC moving at high radial and/or tangential velocities $\sim 50-100$ kpc from the MW. Massive satellites like the LMC leave their mark on the CGM structure of their host galaxies, and signatures from this interaction may manifest in key all-sky observables of the CGM of the MW and other massive galaxies.
\end{abstract}


\section{Introduction} \label{sec:intro}
Galaxies host atmospheres of gas extending out to their virial radii and beyond called the circumgalactic medium (CGM). Observations of the CGM have revealed its structure to be both complex \& multiphase \citep*{Tumlinson2017}. Surrounding the Milky Way (MW), there is evidence of a diffuse, volume-filling CGM component approximately at virial temperatures (T $\sim 10^{6}$ K) and low-densities (n $\sim 10^{-4}$ cm$^{-3}$) traced by highly ionized gas in the X-ray \citep[e.g.][and the review by \citealt{Mathur2022_xray}]{henley_xmm-newton_2012, henley_xmm-newton_2013,2013Miller_hot_halo}. The coldest and densest components of the CGM (T $\sim 10^{4}$ K and n $\sim 10^{-2}$ cm$^{-3}$) are the great complexes of low-ionized and neutral HI clouds, marked by their high velocities ($\gtrsim 200$ km/s) along the line of sight\citep{putman_gaseous_2012, richter_hstcos_2017}. In addition to these reservoirs, intermediate absorbers trace the warm gas (T $\sim 10^{5}$ K), visible in UV absorption spectra of high-redshift quasars and stars in the Galactic halo \citep[e.g.][]{Werk2018}. This component could have several origins, including cooling flows from higher temperature gas and forming at the boundary layers between the cold clouds and the ambient hot medium \citep[e.g.][]{Tumlinson2011, Werk2016}. With its potentially massive baryon budget, the CGM has been theorized to play the role of galactic mediator between the accreting material from the intergalactic medium (IGM) and the interstellar medium (ISM), monitoring the flows of gas in and out of galaxies and regulating their continued star formation \cite[for review see][]{DonahueVoit2022PhR...973....1D}.

However, prior models that have tried to understand the CGM-galaxy connection often neglect the fact that galaxies are not islands, but are in fact, often members of dynamic communities with nearby neighbors. The low gas fractions and diminished star formation rates of satellite galaxies suggest that they are in constant interaction with the CGM of their more massive hosts \citep{Putman2021}. Zoom-in cosmological simulations show that the infall of such satellites supply the CGM of MW-like galaxies ($M_{\rm halo} \sim 10^{12} M_{\odot}$) by $z\sim0$ with $\sim$ 20\% of their mass and metals \citep{hafen_origins_2019}, and perhaps source their cool gas observed at $R > 0.5 R_{\rm 200}$ \citep{Fielding2020,roy_seeding_2023}. Moreover, if those satellites are massive enough themselves, they may possess their own CGM, which would also be colliding with the gaseous halo of their central hosts \citep{Hani2018}. 

Such may be the case in our own Galactic neighborhood, where space-based high precision proper motion measurements of the LMC and SMC, the MW's most massive satellites, suggest that they are on their first passage of the Galaxy \citep{kallivayalil_proper_2006,besla2007,kallivayalil_third-epoch_2013}. These galaxies are also gas-rich, as evidenced by the supply of cold HI gas that both leads and trails the Clouds, and forms a bridge between them \citep[and the review by \citealt{donghia_magellanic_2016}]{Mathewson1974, Putman1998}. In addition to the MC Stream and Bridge, there is an extended reservoir of ionized gas with similar kinematics but (at an assumed distance of 55 kpc) more than 3 times the mass in neutral gas \citep{Bruns2005,fox_cosuves_2014}. This ionized gas may have several different origins, including from the mixing interfaces between the neutral stream and the hot MW halo \citep{fox_exploring_2010} or perhaps even seeded by ram-pressure stripping of the CGM of the LMC (sometimes referred to as the LMC's \enquote{corona}) prior to reaching pericentre  \citep{lucchini_magellanic_2020, lucchini_magellanic_2021, lucchini_properties_2023, krishnarao_observations_2022}. 

A collection of dynamical estimates \citep[see Figure 1 in][]{lucchini_properties_2023,vasiliev_effect_2023} indicate that the LMC is $\gtrsim$ 10\% of the total MW mass \citep{Besla2010}. This means that the infall of the LMC in particular represents a large and recent perturbation to our Galaxy. The total velocity of the LMC has been observed at $\sim 320$ km/s with respect to the MW \citep{kallivayalil_third-epoch_2013,gaia_collaboration_gaia_2021}. Such large speeds exceed the characteristic sound speed expected for $T \sim 10^6$ K gas of $c_s \approx 150$ km/s, implying that the LMC should drive a shock from its collision with the hot phase of the MW CGM. This has been explored in the context of the collision of the LMC's HI disk and the CGM \citep{de_boer_bow-shock_1997,setton_large_2023}, but what remains unexplored is the detailed changes in the physical and kinematic properties of the CGMs of the MW and LMC owing to this supersonic collision. If the LMC indeed hosted it own CGM, then its first infall offers us a prime opportunity to study the interaction between the LMC and the Milky Way and their respective CGM in action.

However, studies on the dynamical response of the MW system to the first infall of the LMC has largely centered on the Galaxy's collisionless components: its dark matter and stellar halo. N-body simulations of the MW and the LMC interaction have found that the LMC generates density wakes and kinematic distortions in the MW’s dark matter and stellar halo, generating an overdense \textit{transient} wake which trails the orbit of the LMC and a global \textit{collective} response \citep{Gomez2015ApJ,garavito-camargo_hunting_2019, garavito-camargo_quantifying_2020, cunningham_quantifying_2020, Peterson2020MNRAS, Foote2023}, from a combination of low-order resonances in the DM halo with the LMC's orbit and the reflex motion of the inner MW halo relative to its outskirts and towards the past pericentre position of the LMC \citep{weinberg_self-gravitating_1989,weinberg_dynamics_1998}. The reflex motion has been detected in the all-sky kinematic patterns of stars in the outer halo, in the form of redshifted and blueshifted radial velocities in the Northern/Southern hemispheres \citep{petersen_detection_2020,Erkal2021, conroy_all-sky_2021,Yaaqib2024MNRAS,Chandra2024}. Such physical and kinematic distortions induced from the gravitational interaction between the MW and LMC should also be shaping the CGM of the MW halo, in addition to its dark matter and stellar components.

In this article, we will explore how the infall of massive satellites imprint themselves on the CGM structure of their host galaxies using idealized, high-resolution simulations of a MW-like CGM with a live halo and an infalling LMC-like satellite on its first passage. We find that the infall of the LMC induces large-scale distortions to the MW CGM's physical and kinematic properties as a consequence of both the dynamical response of the MW halo to the LMC's passage and the halo-scale shock wave from the supersonic collision of their respective gaseous halos. We will then end with a discussion on how the signals from this interaction, in addition to the reservoir of stripped LMC CGM material along its past orbit, manifest in key observables of the MW CGM, and future work exploring the broad consequences for our Galaxy, its surrounding satellites, and the CGM of similar systems.

In the following section, we begin with a brief description of the MW-LMC simulations used in this work and the properties of our assumed orbital trajectory for the LMC (Section \ref{sec:meth}). In Section \ref{sec:results}, we present the following results of our simulation run: the temporal evolution of our fiducial MW-LMC run, changes in the physical properties of the MW CGM as a product of the encounter, an analysis of the resultant shock, the gas response in kinematics, and its sensitivity to the assumed CGM mass of both galaxies. We follow that in Section \ref{sec:disc} with a discussion of the encounter's manifestation in MW CGM observables, its potentially extensive implications for a range of properties of the MW system, and finally, a prediction for the acceleration of cosmic rays in the CGM from the shock. We then close with our summary and conclusions (Section \ref{sec:summ}).

\begin{figure*}[ht!]
    \centering
  \includegraphics[width=1\linewidth]{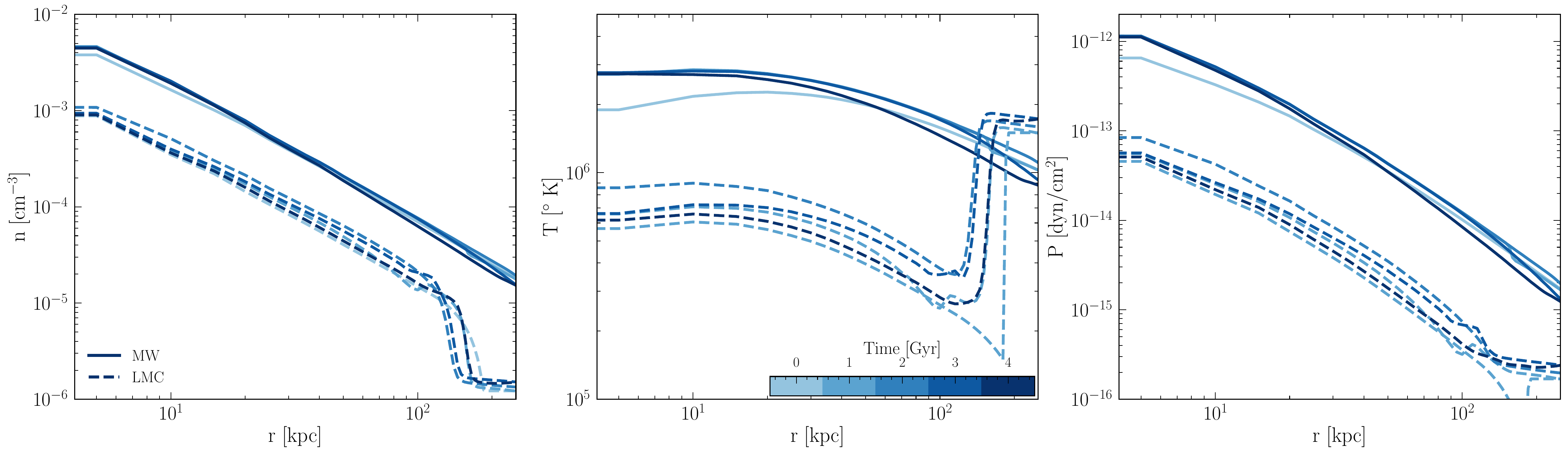}
  \caption{Number density (\textbf{left}), temperature (\textbf{center}), and pressure (\textbf{right}) profiles for the MW CGM (\textbf{solid}) and LMC CGM (\textbf{dashed}) evolved in isolation at times $t = 0,1,2,3,4$ Gyr. Gaseous halos are initialized in approximate hydrostatic equilibrium with live DM halos with the assumed initial parameters listed in Table~\ref{tab: params}. All profiles extend out for either galaxy until they reach a low-density background floor at $\sim 2 R_{\rm 200}$.}
  \label{fig: prof}
\end{figure*}

\section{Methodology} \label{sec:meth}
The simulations described in this work were run with \textit{Enzo}, an Eulerian hydrodynamics code with adapative mesh refinement (AMR) for enhanced spatial and temporal resolution \citep{bryan_enzo_2014}. For our fiducial run, the size of our simulation domain is $L = 3$ Mpc with a root grid resolution of $N_{\rm root}$ = 256 and $l_{\rm max} = 7$ maximum levels of additional refinement. With this resolution, we achieve a minimum grid size of $\Delta x_{\rm min} = L / (N_{\rm root} 2^{l_{\rm max}}) \approx 92$ pc. In addition, we use particles masses of $m_{\rm DM} = 2 \times 10^{5} M_{\odot}$ and $m_{\star} = 2.5 \times 10^{4} M_{\odot}$ to represent the dark matter and stellar components. Since the emphasis of this work is on the large-scale response of the MW CGM to the LMC, we do not explicitly include the interstellar medium of either galaxy, the SMC, radiative cooling, or star formation / feedback in these simulations, and leave these extensions to future work. 

\begin{table}[ht!]
\centering
\begin{tabular}{lll}\hline
Model Parameter &  MW & LMC\\\hline\hline 
 Halo Mass ($M_{\rm 200}$)  &  $1.1 \times 10^{12} M_{\odot}$ & $1.8 \times 10^{11} M_{\odot}$ \\
 Halo Radius ($R_{\rm 200}$)  &  215 kpc & 117 kpc \\
 NFW Scale Radius ($r_s$) &  17 kpc & 13 kpc \\
 Concentration ($c$) &  12 & 9 \\
 Stellar Mass ($M_{\rm \star}$) & $5 \times 10^{10} M_{\odot}$ & $2.5 \times 10^{9} M_{\odot}$\\
 Disc Scale Radius ($r_{\rm d}$) & 3 kpc & 1.7 kpc \\
 N Particles (DM) & 9000000 & 1043939 \\
 N Particles (stars) & 2000000 & 100000 \\\hline
 CGM Mass ($M_{\rm CGM}$) & $2.8 \times 10^{10} M_{\odot}$ & $2.4 \times 10^{9} M_{\odot}$\\
 Core Density ($n_0$) &  0.35 cm$^{-3}$ & 0.18 cm$^{-3}$ \\
 Core Radius ($r_c$) &  0.35 kpc & 0.2 kpc \\
 Density Exponent ($\beta$) &  0.5 & 0.5 \\\hline
\end{tabular}
\caption{\label{tab: params}List of Model Parameters for the initial setup of the MW and LMC. MW DM halo and stellar disc properties were rescaled from the \textbf{m12} model first introduced in \cite{Yu2019}, and we use the same LMC model from \cite{besla_role_2012}. The CGM of both galaxies follow a $\beta$-profile \citep{beta1998} and are initialized in approximate hydrostatic equilibrium with their respective DM halos.}
\end{table}

\subsection{Milky Way and LMC Initial Setup}
Now we will describe the initial conditions of our MW and LMC setup, where the relevant quantities are summarized in Table \ref{tab: params}. We use a modified version of the \textbf{m12} model introduced in \cite{Yu2019}, where we rescaled proportionally the initial particle positions and velocities consistent with the virial theorem to represent a MW model with $M_{\rm 200, MW} = 1.1 \times 10^{12} M_{\odot}$ and a radius $R_{\rm 200, MW} = 215$ kpc, in agreement with MW mass estimates \citep{BH_G2016ARA&A}. The halo's mass and radius are defined where the enclosed density of the halo equals 200 times the critical density of the universe at $z=0$. We assume that the dark matter follows a spherically isotropic NFW profile \citep{NFW1996} with a scale radius of $r_s = 17$ kpc and a halo concentration of $c = 12$. Although we do not include an ISM component to the MW, the stellar disc follows an exponential profile with rotational support, with a total stellar mass of $M_{\rm \star, MW} = 5 \times 10^{10} M_{\odot}$ and disc scale radius of $r_{\rm d, MW} = 3$ kpc.

We use the DM and stellar components of the LMC model introduced in \cite{besla_role_2012}. The LMC DM halo has a total mass of $M_{\rm 200, LMC} = 1.8 \times 10^{11} M_{\odot}$, following a Hernquist profile \citep{Hernquist1990} with a scale radius of $r_{\rm H} = 21.4$ kpc and a halo radius $R_{\rm 200, LMC} = 117$ kpc. The LMC contains an exponential stellar disc of $M_{\rm \star, LMC} = 2.5 \times 10^{9} M_{\odot}$ and scale radius $r_{\rm d, LMC} = 1.7$ kpc.

We initialize the CGM of the MW to be in approximate hydrostatic equilibrium (HSE) with its NFW DM halo. This was done by first assuming that the number density of the gas halo follows a $\beta$-profile \citep{beta1998} of the form:
\begin{equation}
    \begin{aligned}
    n (r) = n_0 \Bigg[ 1 + \Bigg( \frac{r}{r_c} \Bigg)  \Bigg]^{-3\beta /2},
    \end{aligned}
\end{equation}
where $n_0$ and $r_c$ describe the core density and radius respectively and $\beta$ encapsulates the falloff of the density profile at radii greater than $r_c$. We assume $n_0 = 0.35$ cm$^{-3}$, $r_c = 0.35$ kpc, and $\beta = 0.5$, resulting in a total MW CGM mass of $M_{\rm CGM, MW} ( < R_{\rm 200, MW}) \approx 2.8 \times 10^{10} M_{\odot}$. With this density profile, we then derive the pressure and temperature profiles that satisfy the HSE condition:
\begin{equation}
    \begin{aligned}
        \frac{dP_{\rm CGM}}{dr} (< r) = \frac{d\Phi_{\rm DM}}{dr} (< r) \rho_{\rm CGM} (< r)
     \end{aligned}
\end{equation}   
for an ideal gas. 
We repeat the procedure for the LMC CGM, but we assume different parameters for its realization. For simplicity, we assume that the LMC CGM also follows a $\beta$-profile with $\beta = 0.5$, but that its core radius is reduced by a factor $\sim R_{\rm 200, LMC} / R_{\rm 200, MW}$. We assume for our fiducial runs a core density that results in a total LMC CGM mass of $M_{\rm CGM, LMC} ( < R_{\rm 200, LMC}) \approx 2.4 \times 10^{9} M_{\odot}$, a CGM mass ratio of roughly $1:10$ with the MW CGM. We note that the total mass of the LMC CGM prior to infall and its mass ratio with the MW CGM is highly unconstrained, and so we assumed a CGM mass ratio that is somewhere in between the DM mass ratio ($\sim 1:6$) and the stellar mass ratio ($\sim 1:20$) between the two galaxies as our starting point. In section \ref{sec: vary}, we explore the sensitivity of our results to this assumed initial CGM mass ratio.  

In Figure \ref{fig: prof}, we plot the density and derived temperature and pressure profiles from HSE for the MW and LMC CGM at $t=0$, and at times $t=1,2,3,4$ Gyr when evolved in isolation. We find that our assumed densities profile are roughly stable over 4 Gyr of simulated time. At 50 kpc, this yields a MW CGM number density of $n (r = 50$ kpc) $\approx 2 \times 10^{-4}$ cm$^{-3}$, broadly consistent with existing observational constraints \citep[see Figure 3 from][]{Voit2019}. Since the gas profiles were initialized with respect to the DM halo, the inner halo of both galaxies rapidly compresses in response to the disc potential, increasing the density and pressure profiles over time and causing the inner temperature profile to fluctuate within a factor of 2. Overall, we find that both halos produce radial profiles that are relatively stable for 4 Gyr of simulated time. 

\begin{figure}[h!]
    \centering
  \includegraphics[width=1\linewidth]{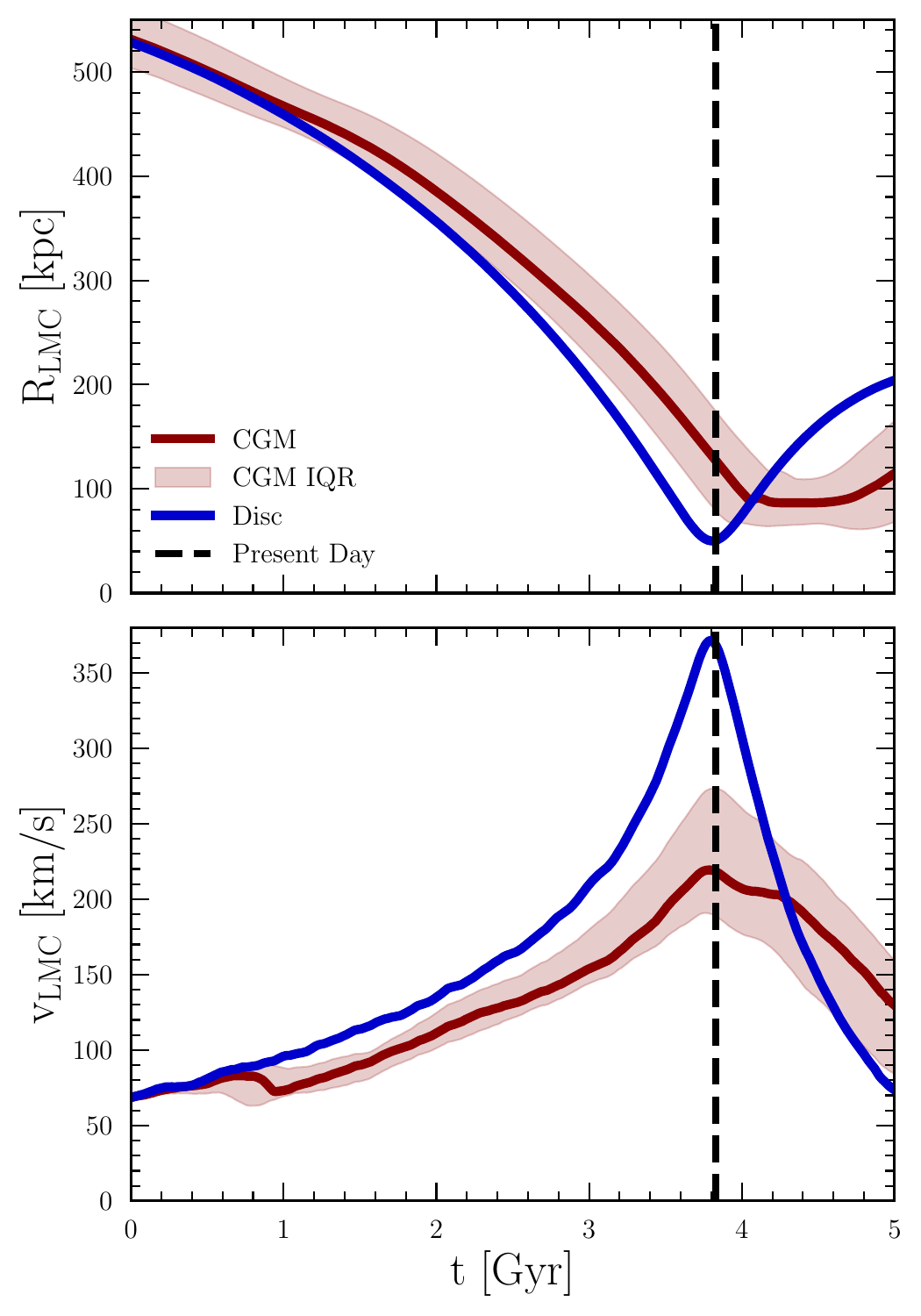}
  \caption{Galactocentric radius (\textbf{upper}) and total velocity (\textbf{lower}) of different components of the LMC over the course of the simulation. The blue line in both plots traces the average radius and velocity of the LMC stellar disc with respect to the Galactic center. The solid red line marks the mass-weighted median radius and velocity of LMC CGM gas, and the shaded region marks the weighted 25th and 75th percentile range of both quantities respectively.}
  \label{fig: orbit}
\end{figure}

\begin{table*}
\begin{tabular}{llll}\hline
 &  Initial (t=0 Gyr) & \enquote{Present-Day} (t=3.83 Gyr) & Observed \citep{kallivayalil_third-epoch_2013} \\\hline\hline 
 Position (kpc) &  (63.97, 524.42, -4.05) & (-0.17, -43.05, -26.48) & (-0.8, -41.5, -26.9) \\
 Velocity (km/s) &  (2.14, -54.72, -41.42) & (-83.27, -223.55, 280.64) & (-57$\pm$12, -226$\pm$11, 252$\pm$16) \\\hline
\end{tabular}
\caption{\label{tab: orbit} Galactocentric positions and velocities for the LMC, defined using the average position of its stellar disc, at its initial conditions (t=0 Gyr) and present-day position from the simulation (t=3.83 Gyr), alongside the observed values of its current position and velocity from \cite{kallivayalil_third-epoch_2013}.}
\end{table*}

\subsection{LMC Orbital Trajectory}
The orbit of the LMC is based off a modification of the Model 1 orbit trajectory first introduced in \cite{besla_role_2012}. We made two modifications to these initial Galactocentric positions and velocities: (1) we backward-integrated the LMC's position from its initial position in \cite{besla_role_2012} to a larger Galactocentric radii ($R_{\rm LMC} = 528$ kpc $\sim 2 R_{\rm 200, MW}$) using the orbit integration code \textit{Gala} \citep{gala}, and (2) varied the initial velocity components until it produced an orbit closest to the observational constraints from \cite{kallivayalil_third-epoch_2013}. The former was done such that the DM and gaseous halos of both galaxies could be initialized in relative isolation before the interaction begins.

In Figure \ref{fig: orbit}, we plot the Galactocentric radii and total velocities for the LMC disc and CGM as a function of time. The LMC disc values are from the average position and velocity of its constituent particles. The CGM quantities are the mass-weighted median of the position and velocity. We see that for the first $\sim$1 Gyr, the position and velocities of the stellar and CGM components move in lockstep as the LMC accelerates in the outskirts of the MW potential where the ram-pressure from the ambient density is insufficient to uncouple the LMC CGM's gas motion from its halo. 

However this begins to change once the LMC approaches $\sim R_{\rm 200,MW}$. As the LMC descends deeper into the MW's potential well and approaches pericentre, the greater velocities of the LMC combined with the climbing densities of the MW CGM enhance the ram-pressure experienced by the LMC CGM, slowing its overall velocity with respect to the disc. This separation in velocity is largest once the LMC reaches pericentre at $r_{\rm d, peri} = 50$ kpc where its velocity peaks at $v_{\rm d, peri} = 372$ km/s and the median LMC CGM velocity peaks at $v_{\rm CGM, peri} = 219$ km/s. After the disc's pericentre, the disc and CGM seem to be moving with similar total velocities but with different behavior in their positions. The disc moves past its closest approach to larger radii, but the bulk of the gas hovers at a radius around 100 kpc, seemingly disconnected from the motion of the LMC disc.

In Table \ref{tab: orbit}, we present our initial Galactocentric positions and velocities for the LMC's disc component, alongside the simulation estimates at t=3.83 Gyr, the snapshot in time where we achieve our closest match to the observational constraints. We note that, given our simplified MW potential and lack of an SMC companion, our orbit can only be an approximation of the real LMC's past trajectory. Despite this limitation, we are able to get the LMC within $\sim 2 \sigma$ of the measured LMC's Galactocentric positions and velocities. As we will demonstrate in the next section, the emphasis of this work on the large-scale response of the MW CGM to the infall of the LMC and the collision of their respective gaseous halos is most sensitive to the total velocity of the LMC, due to its ability to drive large-scale shocks, and we do not expect our results to change significantly with the use of other first-infall orbit models consistent with observed proper motions.

\section{Results} \label{sec:results}

After describing the different aspects of our simulation, we now present a brief description of the time evolution of the MW-LMC encounter, the response of the MW CGM to the LMC's infall seen in its physical properties, and characterize the properties of the shock front produced from its supersonic motion. We will then focus on the kinematic imprints left on the CGM gas of the respective galaxies. 

\begin{figure*}[ht!]
    \centering
  \includegraphics[width=1\linewidth]{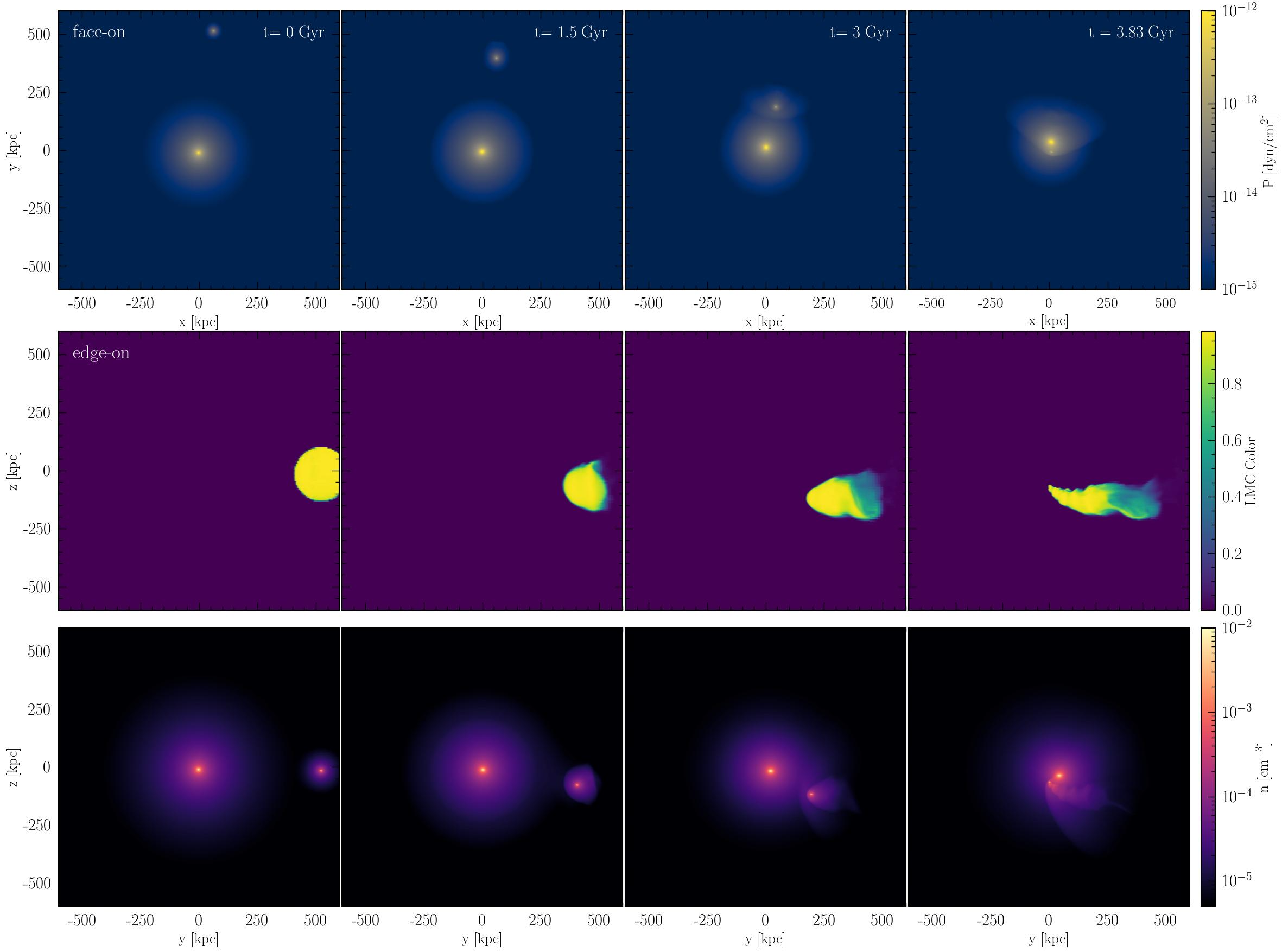}
  \caption{Evolution of the MW-LMC encounter at times $t=0,1.5,3,3.83$ Gyr. Each row depicts the projected gas pressure (\textbf{top}) \enquote{face-on} in the x-y simulation plane, the LMC Color (\textbf{middle}), and the projected CGM number density \enquote{edge-on} in the simulation y-z plane (\textbf{bottom}). The LMC CGM feels the headwind from the MW CGM, unbinding most of it over time. During infall, the colliding LMC CGM generates a large-scale shock across the MW halo, sharply visible in pressure and density. Gas belonging to the LMC CGM is initialized with a \enquote{LMC Color}, a passive scalar quantity equal to 1 for all LMC CGM gas and O for all other gas at t=0. At later times, this LMC Color is diluted in the LMC CGM from mixing with the MW CGM, dropping below 1 in the stripped tail.} 
  \label{fig: tevol}
\end{figure*}

\subsection{Evolution of the MW-LMC CGM Interaction} \label{subsec:evol}

First, we describe the evolution of the MW-LMC interaction and its impact on the CGM of both galaxies through time. Figure \ref{fig: tevol} shows the projected gas pressure "face-on" (w.r.t. the MW disc) in the simulation x-y plane in the top panel, and in the middle and bottom panels, displays the \enquote{LMC Color} and the projected number density edge-on in the y-z plane. The LMC Color is defined as a passive scalar that marks gas originating in the LMC satellite. All CGM gas belonging to the LMC is initialized with a color of 1 (color of 0 for MW and background gas), and this color gets diluted as gas mixes between the two reservoirs. The columns from left to right advance in time and depict the interaction during four different regimes.

At t=0, the system reflects the initial conditions, where the centers of both gaseous halos are separated by $528$ kpc and are in approximate HSE with their respective dark matter halos. As the interaction ensues, the outskirts of both halos are increasingly affected, reflected in the middle-left column at t=1.5 Gyr. During this time, ram-pressure from the low-density gas beyond the MW virial radius starts to sculpt the morphology of gas in the LMC's outskirts, breaking its spherical symmetry around the disc (disc not plotted explicitly but tracks with the dense core of LMC CGM). By t=3 Gyr, after the LMC has passed within $\sim R_{\rm 200, MW}$, ram-pressure from the MW CGM blows away almost all gas leading the LMC, forming a \enquote{head-tail} morphology where the stripped LMC gas trails the disc along its past orbit. This regime captures two other notable developments. First, the LMC color tracer starts to decline at the far end of the trailing gas. This marks the gas that was stripped early during the interaction and is now mixing with the MW's hot halo. The other feature worth emphasizing is the large-scale compressive front gaining prominence in the MW CGM, boosting the gas pressure and density in the region leading the LMC. In the last column at t=3.83 Gyr, after the LMC passes pericentre and nears its present-day position, the compressive front grows steeper, forming a merger-induced shock wave with a length size on the order of the radius of the MW halo. By this time, gas that initially fell in with the LMC has been stripped, but most of the gas that has been recently stripped remains unmixed with the MW hot halo. This global response of the MW CGM to the infall of the LMC leads to significant differences compared to the MW in isolation, as we explore in the next section.

\begin{figure*}[ht!]
    \centering
  \includegraphics[width=1\linewidth]{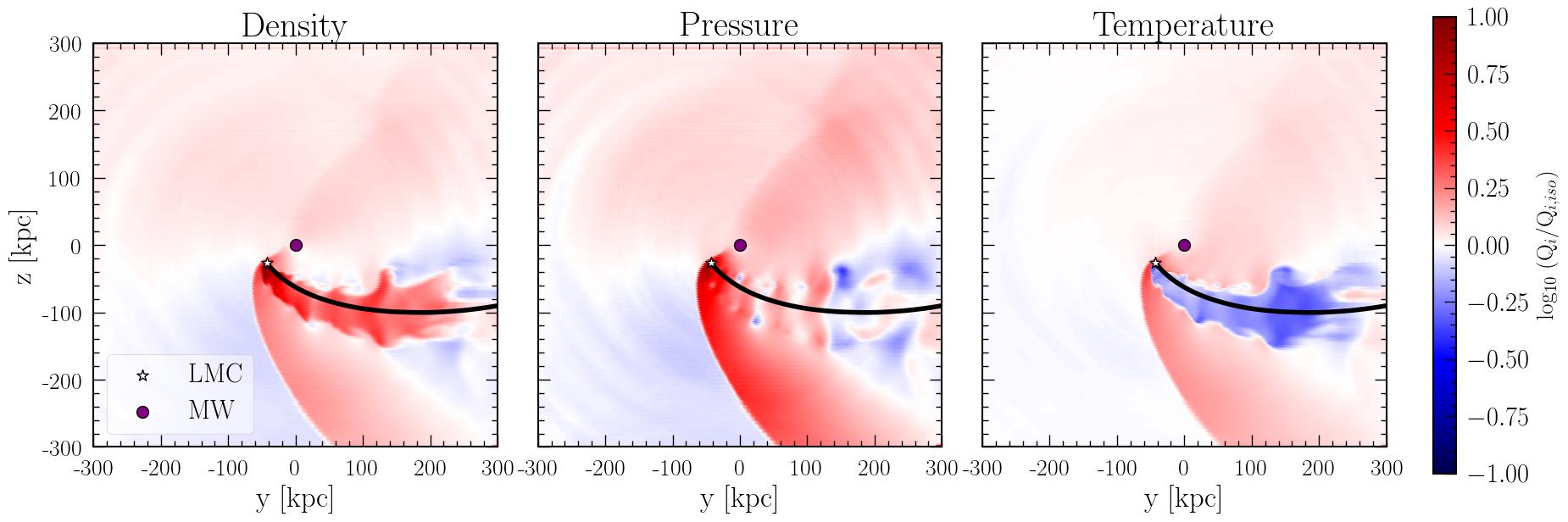}
  \caption{Changes in CGM density (\textbf{left}), pressure (\textbf{center}), and temperature (\textbf{right}) in the Galactocentric y-z plane, in approximate alignment with the LMC's orbit. These are slice plots with a thickness of 2.3 kpc centered on the MW disc. The colormap shows the log of the ratio of each CGM quantity, log$_{\rm 10}$ (Q$_i$/Q$_{i,\rm iso}$), between its value in the simulation with an infalling LMC closest to its observed Galactocentric position ($t=3.83$ Gyr) and the value of that same quantity from an isolated MW CGM at the same time. Prominent features across all three maps include the stripped material of the LMC CGM along its past orbit, the shock front from the supersonic collision of the gaseous halos, and the dipole-like collective response above and below the Galaxy from the reflex motion of the MW towards the pericentre position on the LMC. The white star and the black line marks the current position and past orbital trajectory of the LMC respectively. The LMC CGM-MW CGM interaction is predicted to create large scale distortions in the physical properties of the gas presently surrounding the MW.}
  \label{fig: res}
\end{figure*}

\newpage
\subsection{Impact on the CGM Physical Properties} \label{subsec:res}

The infall of the LMC imprints itself on the physical properties of the MW CGM. In Figure \ref{fig: res}, we display the pattern and scale of the MW CGM response in density, pressure, and temperature in a slice across the Galactocentric y-z plane at t=3.83 Gyr, our assumed present-day snapshot. We quantify this change by taking the log of the ratio of a given quantity, such as density, pressure, or temperature, from the CGM in the MW-LMC simulation and its corresponding value at the same position with respect to the Galaxy in an isolated MW simulation with no LMC: $\Delta Q_i \equiv \log_{10} (Q_{i} / Q_{i,\rm iso}$). These maps allow for an easy identification of the prominent features of the MW-LMC interaction and the relative amplitudes of each signal. 


In each panel of Figure \ref{fig: res}, we see similar global features across the CGM. The feature with the largest enhancement in density tracks in lockstep with the past orbital trajectory of the LMC and represents the contribution from stripped LMC CGM gas due to ram-pressure from the MW atmosphere. The stripped LMC gas leaves a near continuous stream of trailing gas stretching out to $\sim 300$ kpc in length in the y-z plane, with a density about 3 times greater ($\Delta \rho \approx 0.5$) than the background MW CGM density. This density contrast grows larger closer to the LMC. We see this stripped LMC gas prominently in temperature as well, reflecting in part the initial difference in virial temperatures between the two gas halos. Shear flows at the interface between the stripped gas and the MW CGM give rise to the Kelvin-Helmholtz (KH) instability, generating large turbulent eddies along the stream as the two gas reservoirs begin to mix. We see analogous features in the pressure map, where these eddies generate local fluctuations in gas pressure along the interface with MW gas. Since we do not include the cold gas of the LMC (or the SMC), there is no neutral-HI MC Stream, however parts of the Stream would likely coincide with this stripped ionized gas from the LMC CGM. In summary, the MW-LMC interaction leaves compressed, cool LMC CGM gas along the satellite's past orbital trajectory that is actively mixing with the hot halo by the present-day. This gas is likely an important contributor to the observations of ionized gas associated with the MC Stream.

Just as striking as the stripped gas is the sharp jump in density, pressure, and temperature along the curve of the wave front leading the LMC. This compressive front upstream of the LMC is the merger-induced shock from the supersonic collision between the respective CGM of the MW and the LMC. This shock front is similar to the bow shock described in \cite{setton_large_2023}, but while their work described the shock expected from the collision between the LMC ISM and the MW CGM, our work emphasises the shock produced at the CGM-CGM interface. The shape and complex morphology of the shock is a product of the overall encounter. The large surface area for the collision due to the LMC CGM's inclusion compared to the ISM alone, and a wide Mach angle, produce a shock front that extends beyond the radius of the MW halo.  The LMC and its retained gas actively drive its shock below the MW, but a fainter jump is also visible above the Galaxy, sharing a cross-section with the MW disc. This segment above the MW is likely a remnant from an earlier phase of the shock before pericentre when the shock was weaker and more symmetrical around the LMC, but this portion of the shock weakened compared to the counterpart below the MW as it climbed the steep pressure gradient of the inner CGM. As we detail in section \ref{subsec:shock}, the enhancement across the shock in density, pressure, and temperature compared to the isolated MW are consistent with a shock of Mach number $\mathcal{M} \approx 2$. 

Another global feature seen in the residual maps is the global dipole-like signal, characterized by an enhancement (diminution) in density, pressure, and temperature that maps onto the Northern (Southern) hemisphere of the MW. This collective response is due to the reflex motion of the inner halo towards the past pericentre position of the LMC and away from the outer halo in the Northern hemisphere, resulting in a halo-scale distortion centered on the Galaxy. Due to the radial dependence of the effect, the collective response in the CGM is most readily seen in quantities that have steep radial gradients. This explains why the collective response in density and pressure, quantities where we assume a radial gradient, are characterized by an over(under)-enhancement of $\Delta \rho$ $\sim$ $\Delta P$ $\sim$ $\pm 0.1-0.2$, while the signal in temperature, where the radial gradient is flatter, is weaker. 

In summary, the passage of the LMC produces order-unity distortions to the global structure of the Galaxy’s CGM from a combination of stripped LMC CGM gas along the past orbital trajectory, merger-induced shocks from the CGM-CGM collision, and the differential dynamical response of the Milky Way halo to the LMC's infall.

\begin{figure}[ht!]
    \centering
  \includegraphics[width=1\linewidth]{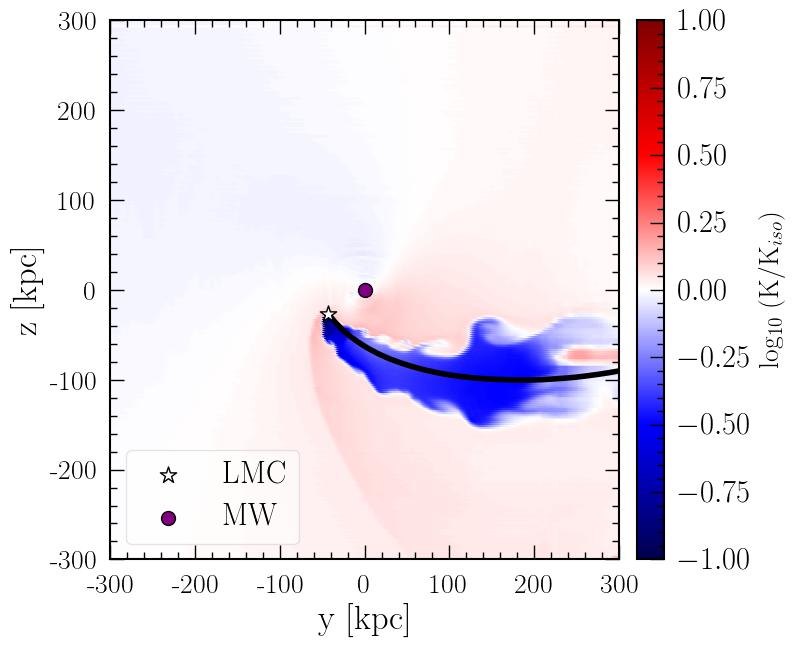}
  \caption{Similiar to Figure \ref{fig: res}, but for changes in CGM entropy. The color map is changed to show the log of the ratio of the CGM entropy, log$_{\rm 10}$ (K$_i$/K$_{i,\rm iso}$), between its value in the simulation with an infalling LMC closest to its observed Galactocentric position ($t=3.83$ Gyr) and the entropy from the isolated MW CGM run. The shock emerges as a sharp discontinuity in the entropy along the curvature of the wave front. The low entropy region reflects the cool, stripped material from the LMC CGM.} 
  \label{fig: resK}
\end{figure}




\subsection{Characterizing the Shock Front \label{subsec:shock}}
Under ideal conditions, the pre- and post-shock regions in the rest frame of the shock are related by the familiar Rankine–Hugoniot (RH) jump conditions and depend on
the Mach number of the shock, defined as the ratio of the velocity in the rest frame of the shock and the gas sound speed $\mathcal{M}_1 \equiv v_1 / c_s$. The jump in entropy across the shock is defined as,

\begin{equation}
    \begin{aligned}
         \frac{K_1}{K_0} = \Bigg(\frac{5\mathcal{M}_1^2 - 1}{4} \Bigg) \Bigg(\frac{4 \mathcal{M}_1^2}{\mathcal{M}_1^2 + 3}\Bigg)^{-5/3} .
    \end{aligned}
\end{equation}

In Figure \ref{fig: resK}, we display the difference in entropy between our MW CGM and MW-LMC CGM simulation, just as we did for density,  pressure, and temperature in the prior section. The stripped gas of the LMC is clearly visible and its low entropy stands out compared to the background of the MW hot halo. We can also see the faint signal from the collective response, but in the case of entropy, we see a subtle enhancement in the entropy, on the order of $\Delta K \sim 0.05$ in the Southern hemisphere and a comparable diminution in the Northern hemisphere, the reverse of what is seen in density and pressure. This is due to entropy being an increasing function of Galactocentric radius, whereas both density and pressure are assumed to decrease with radius. 

We see a sharp jump in the entropy that traces the shock front. If we use $\Delta Q_{i}$ as our ratio between the pre- and post-shock regions, then the largest entropy jump just ahead of the LMC reaches $10^{\Delta K} \backsimeq 1.42$. When we evaluate the jump in the other CGM properties at the same location, we find ratios of the pre-shock to post-shock conditions of $10^{\Delta Q_i} \backsimeq 1.98$, $4.46$, and $2.25$ for the density, pressure, and temperature respectively. When we compare these jumps to the predicted enhancements from the RH jump conditions, we find that they are consistent with a $\mathcal{M} \approx 2$ shock. For a sound speed of $c_{s} \approx 150$ km/s, gas moving at $\sim 300$ km/s is required for a $\mathcal{M}=2$ shock, which is consistent with the peak of typical velocities for the LMC CGM gas at this time (see Figure \ref{fig: orbit}).

Next we measure the stand-off radius of the shock, $R_{\rm so}$, defined as the separation distance between the center of the LMC and the shock front. We define a ray starting from the LMC and point it along the direction of the LMC's velocity vector, and mark $R_{\rm so}$ by the location of the sharp discontinuities from the shock in density, pressure, and temperature along the ray. By this method, we measure a stand-off radius of $R_{\rm so} = 7.4$ kpc, slightly larger than the stand-off radius of $R_{\rm so} = 6.7$ kpc measured in the LMC ISM wind-tunnel simulations of \cite{setton_large_2023}. The stand-off radius in the ideal scenario of rigid objects is a function of both the radius of curvature of the object and its Mach number \citep{Billig1967}. From this intuition, we would expect a larger $R_{\rm so}$ at fixed Mach number when we include the LMC CGM compared to the ISM alone due to the larger radius of LMC gas participating in the supersonic collision, which is what we find. 

\begin{figure*}[ht!]
    \centering
  \includegraphics[width=0.8\linewidth]{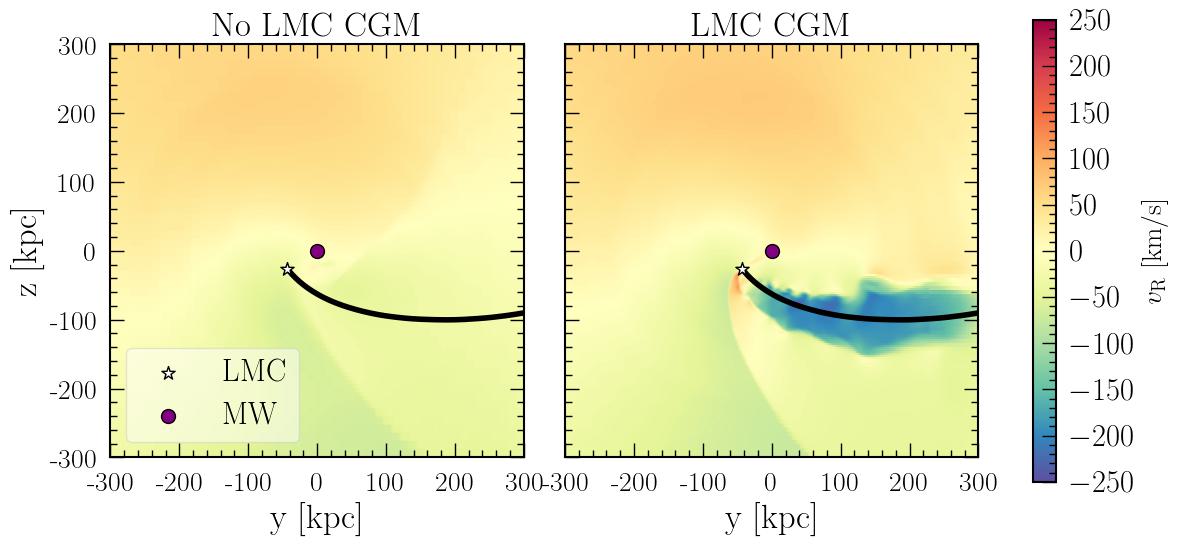}
  \caption{Radial velocities of CGM gas in the Galactocentric y-z plane from two MW-LMC runs: one where the infalling LMC has its own CGM (\textbf{right}) and one without an LMC CGM (\textbf{left}). The white star and the black line marks the current position and past orbital trajectory of the LMC respectively. The No LMC CGM case emphasizes the dipole-like response in radial velocities from the reflex motion of the MW. Including the LMC CGM produces shock-accelerated MW CGM gas with positive radial velocities and a long tail of stripped material moving at $\lesssim -200$ km/s radial velocities along the LMC's past orbit. The bulk motion from systematic velocities of the MW during the interaction have been subtracted.} 
  \label{fig: vR}
\end{figure*}

\subsection{Response in Gas Kinematics} \label{sec: vels}
In the prior sections, we saw that the infall of the LMC produces large-scale changes in CGM density, temperature, pressure, and entropy. Now we consider the gas kinematic response, and demonstrate how the kinematic picture of the gas compliments the prior analysis in the CGM's physical properties. 

\subsubsection{Radial Motions}
In Figure \ref{fig: vR}, we display the kinematic response of the MW CGM in radial velocity with respect to the Milky Way center in the y-z plane from two MW-LMC runs: our fiducial run with an LMC CGM (right) and one without an LMC CGM (left). This comparison allows us to disentangle the superimposed signals from collisional hydrodynamic effects and purely gravitational processes. On the right, we see that including the CGM produces two prominent features in the radial motions of the Southern hemisphere: (1) a stream of trailing material marked by its large negative radial velocities and (2) shock-heated MW CGM gas accelerated to comparable radial velocities to that of the LMC. For the former, this is the same gas identified earlier as the stripped CGM gas of the LMC. We see that most of this trailing gas has notably large negative radial velocities with respect to the rest of the CGM in the range of $v_{\rm R}$ $\sim$ -$150-200$ km/s and lies roughly 100 kpc from the Galaxy. LMC CGM gas that has been more recently stripped has positive radial velocities closer to that of the LMC disc itself. 

In addition to gas associated with the LMC, in the right panel of Figure \ref{fig: vR} we see the sharp outline of the shock. MW CGM gas sweeping through the shock is accelerated to positive radial velocities. The strongest jump occurs in the vicinity of the LMC, where MW CGM reaches radial velocities of $v_{\rm R}$ $\sim$ $100$ km/s. This demonstrates a stark contrast in the kinematic signature of gas associated with the MW and the LMC that persists to the present-day, a distinction that is reflected in their physical properties. In the left panel, when the LMC-CGM gas is not included, we see that the LMC still drives a wave in the MW CGM from its dynamical interaction, but this wave is not strong enough to produce a shock. 

We also see in both scenarios that the presence of the LMC shifts the barycenter of the MW-LMC system and accelerates the MW towards the past orbital position of the LMC at pericentre. This reflex motion of the MW with respect to the halo, responsible for the collective response seen in the CGM physical properties, manifests as a globe dipole-like signal in CGM radial velocities, increasing (decreasing) velocities in the Northern (Southern) hemisphere. Because this effect is caused by the differential response of the MW and the outer halo from the infall of the LMC, we see in the case where the LMC does not have a CGM, the reflex motion signal peaks around $R_{\rm 200, MW}$ at $v_{\rm R}$ $\sim$ $\pm 70$ km/s, but at smaller radii, we still see $v_R$ $\sim$ $\pm 30-50$ km/s from 50 to 100 kpc away from the disc. This is consistent with the radial velocity measurements of stars in the halo, shifted by this same global reflex motion \citep{petersen_detection_2020,Erkal2021,Chandra2024,yaaqib_radial_2024}

In summary, the radial motions of the CGM are shaped by a dynamical response of the MW to the LMC's infall, creating a global dipole-like response in radial velocities, and a hydrodynamical response from a combination of stripped LMC CGM gas with large negative radial velocities in the Southern Hemisphere that trace the past orbit of the LMC and shock-heated MW CGM gas accelerating with the LMC. 

\begin{figure*}[ht!]
    \centering
  \includegraphics[width=0.8\linewidth]{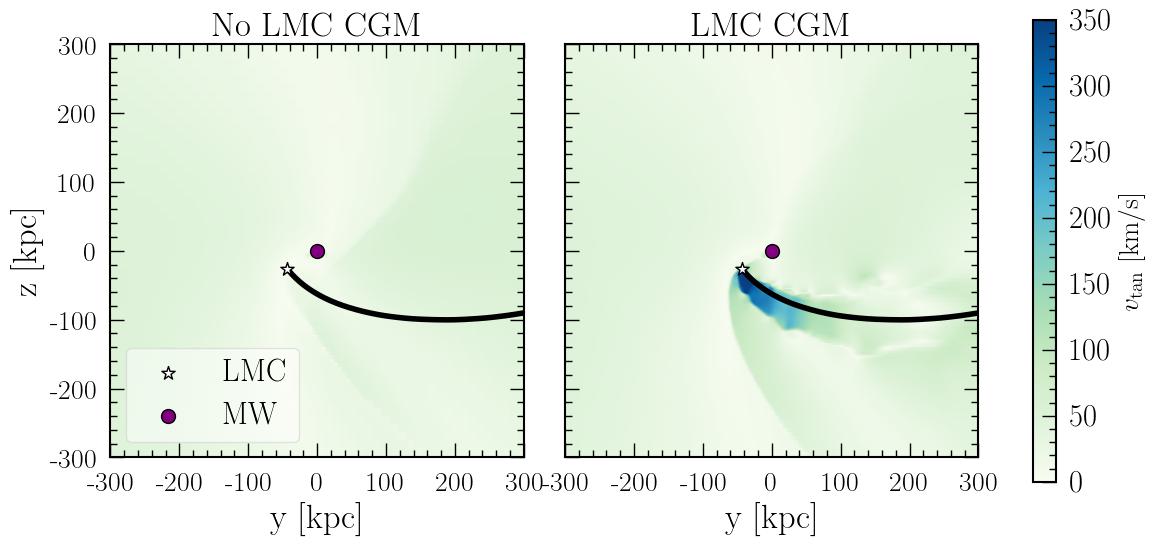}
  \caption{Tangential velocities of CGM gas in the Galactocentric y-z plane from two MW-LMC runs: one where the infalling LMC has its own CGM (\textbf{right}) and one without an LMC CGM (\textbf{left}). The white star and the black line marks the current position and past orbital trajectory of the LMC respectively. In the No LMC CGM case, the LMC modestly accelerates MW CGM gas in its dynamical friction wake. Including the LMC CGM, the stripped material is characterized by large tangential velocities, reaching velocities $>$ 300 km/s in the recently stripped material. Similar to Figure \ref{fig: vR}, the bulk motion from systematic velocities of the MW during the interaction have been subtracted.} 
  \label{fig: vtan}
\end{figure*}

\subsubsection{Tangential Motions}
Figure \ref{fig: vtan} shows the gas response of the MW-LMC runs with (right) and without (left) a LMC CGM in tangential velocity. When we include the LMC CGM, we see two distinct features from stripped gas and the shock, similar to that seen in radial motions. The stripped material is characterized by large tangential velocities, exceeding $200$ km/s in the trailing gas and reaching tangential velocities far greater than $300$ km/s in the recently stripped material. These values are in line with the measured tangential velocities of the disc of around $v_{\rm tan}$ $\approx$ 314 km/s \citep{kallivayalil_third-epoch_2013}. 

The shocked gas of the MW CGM is also accelerated in the tangential direction to $v_{\rm tan} \sim 150$ km/s in the vicinity of the LMC and then decreases moving along the curvature of the shock front. In the left panel, for the run with no LMC CGM to drive a shock, we see a much smaller increase in the tangential velocity component of gas caught in the LMC's dynamical friction wake, reaching as high as $\sim 50$ km/s, i.e., subsonic. 

The large tangential and radial velocity components of the trailing gas along the past orbit of the LMC gives us insight on the distribution of its future trajectories. We expect the components closest to the LMC, with large tangential and positive radial velocities (including recently stripped LMC gas \textit{and} shocked MW CGM gas) to follow the LMC as it moves towards apocentre, until ram-pressure and mixing inevitably comes to dominate the gas dynamics. Gas stripped earlier in the LMC's orbit with more negative radial velocities and smaller angular momenta will fall towards the Milky Way on $\sim$Gyr timescales, eventually evaporating into the hot halo on its descent or accreting onto the ISM \citep{fox_exploring_2010}.

\begin{table*}
\centering
\begin{tabular}{llllllll}\hline
 Model & $M_{\rm CGM, MW}$ [$M_{\odot}$] & $M_{\rm CGM, LMC}$ [$M_{\odot}$] & $M_{\rm CGM, LMC}$ ($<$ 20 kpc) & $M_{\rm CGM, LMC}$ (20-250 kpc) &  $R_{\rm so}$ [kpc] \\ 
  time & 0 Gyr & 0 Gyr & 3.83 Gyr & 3.83 Gyr &3.83 Gyr \\\hline\hline 
 Fiducial &  $2.8 \times 10^{10}$ & $2.4 \times 10^{9}$ & $9.5 \times 10^{7}$ (4\%) & $1.8 \times 10^{9}$ (75\%) & 7.4 \\\hline
 Low Mass LMC CGM &  $2.8 \times 10^{10}$ & $1 \times 10^{9}$ & $1.5 \times 10^{7}$ (1.5\%) & $4.5 \times 10^{8}$ (45\%) & 0.9\\
 High Mass LMC CGM &  $2.8 \times 10^{10}$ & $5.7 \times 10^{9}$ & $4.3 \times 10^{8}$ (7.5\%) & $4.6 \times 10^{9}$ (81\%) & 18.2 \\\hline
 Low Mass MW CGM &  $1.2 \times 10^{10}$ & $2.4 \times 10^{9}$ & $2.1 \times 10^{8}$ (9\%) & $1.9 \times 10^{9}$ (79\%) & 24.8  \\
 High Mass MW CGM &  $6.7 \times 10^{10}$ & $2.4 \times 10^{9}$ & $3.7 \times 10^{7}$ (1.5\%) & $6 \times 10^{8} $ (25\%) & 2.3 \\\hline
\end{tabular}
\caption{\label{tab: vary} List of models where we vary the assumed initial CGM mass of the MW and LMC at t=0 Gyr, compared to their values from the fiducial scenario. CGM masses of the other system are held fixed to their fiducial value in each respective Low/High Mass model. We include the measured mass of the LMC CGM within 20 kpc of the center of the LMC disc, as well as the mass of stripped LMC CGM material in the trailing stream 20 to 250 kpc from the LMC at t=3.83 Gyr, alongside their respective percentages of the initial LMC CGM mass. We also list the measured stand-off radius of the shock produced in each run.}
\end{table*}

\subsection{Sensitivity to CGM Assumptions} \label{sec: vary}
Now we will consider how the measured MW CGM response is sensitive to our starting assumptions of the initial MW and LMC CGM masses. We will first explore how varying the mass for the LMC CGM affects the overall response, followed by a similar procedure for the MW CGM. Other assumed aspects of our model, such as the underlying dark matter distribution, are kept consistent with the fiducial runs. Varying the MW/LMC CGM mass does slightly alter the resulting LMC orbit due to changes in the hydrodynamic drag and dynamical friction, but these effects on the LMC orbit appear minor and all comparisons are made at the same present-day snapshot $t = 3.83$ Gyr. To lower the computational expense of these additional runs, we change our root grid resolution to $N_{\rm root} = 128$ and use 6 levels of additional refinement. Experimental runs with our fiducial setup at low resolution show that our results are only weakly dependent on resolution and the same is true for comparable runs with different model assumptions. 

\begin{figure*}[ht!]
    \centering
  \includegraphics[width=1\linewidth]{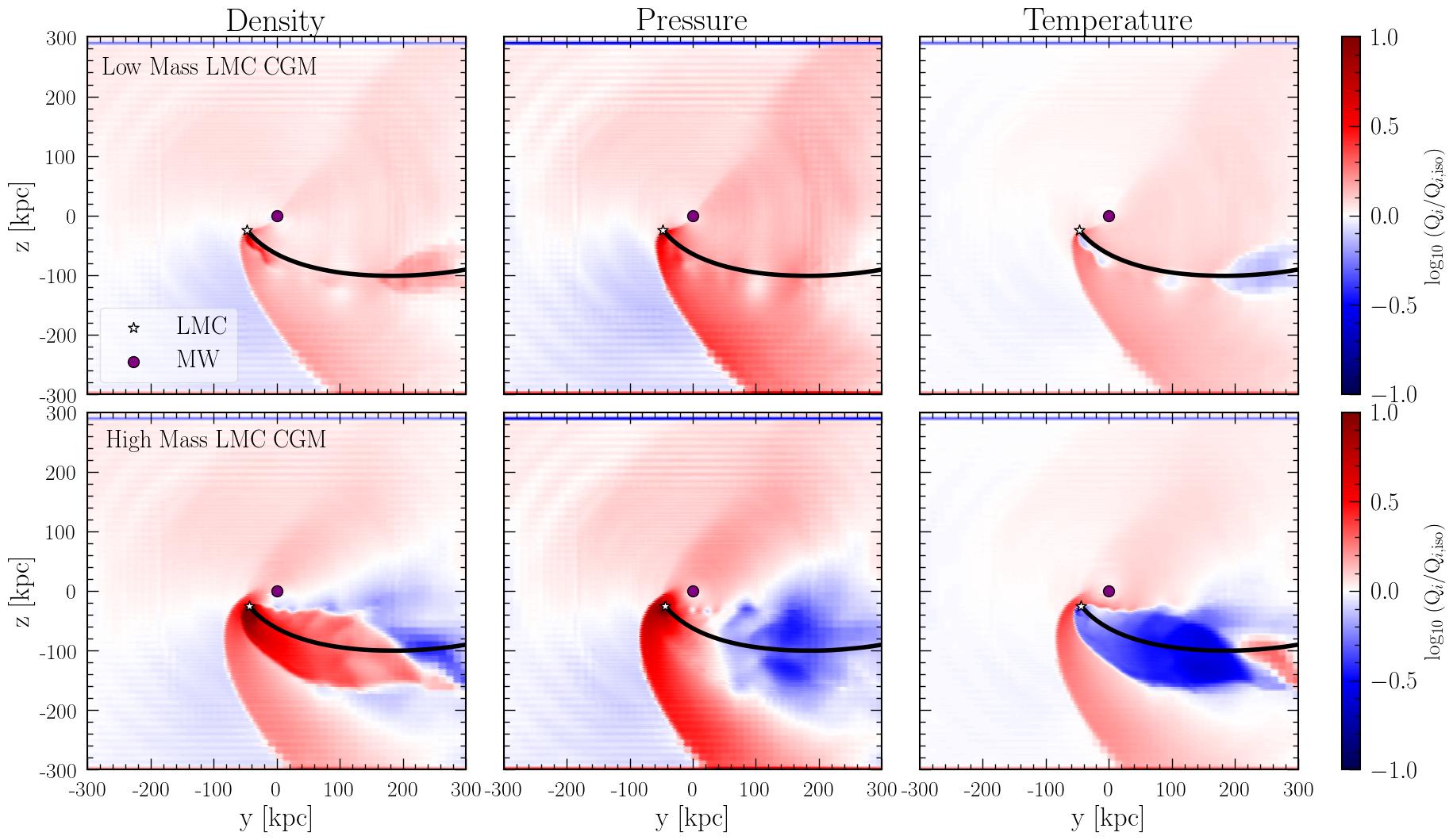}
  \caption{Impact of changing the assumed initial LMC CGM mass. Similar to Figures \ref{fig: res} and \ref{fig: resMW}, but for changes in CGM density (\textbf{left}), pressure (\textbf{center}), and temperature (\textbf{right}) in the Galactocentric y-z plane for both the High Mass (\textbf{bottom}; $M_{\rm CGM,LMC} = 5.7 \times 10^9$ $M_{\odot}$) and Low Mass (\textbf{top}; $M_{\rm CGM,LMC} = 1 \times 10^9$ $M_{\odot}$) models of the LMC CGM. The white star and the black line marks current and past orbital trajectory of the LMC respectively from each respective run to t=3.83 Gyr in simulation time.} 
  \label{fig: resLMC}
\end{figure*}

\subsubsection{Varing LMC CGM Mass}
First we consider how the CGM response depends on the assumed initial mass of the LMC CGM. In Figure \ref{fig: resLMC}, we plot the same residual maps of the change in CGM density, temperature, and pressure with respect to the isolated MW CGM, as was done for our fiducial runs (see Figure \ref{fig: res}), but for a High Mass LMC CGM ($M_{\rm CGM, LMC} = 5.7 \times 10^9 M_{\odot}$) in the bottom row and a Low Mass LMC CGM in the top row ($M_{\rm CGM, LMC} = 1 \times 10^9 M_{\odot}$). We note that what is deemed a \enquote{High Mass} and \enquote{Low Mass} model is somewhat arbitrary compared to the range of conceivable LMC CGM masses, however our choices for the High and Low cases assumed here probe different areas of parameter space where the assumed CGM mass ratios are greater than the DM mass ratio ($\sim$ 1:5 for High Mass) and less than the stellar mass ratio ($\sim$ 1:27 for Low Mass). The assumed initial masses of the MW/LMC CGM used in each model, as well as their fiducial values and present day properties, are listed in Table \ref{tab: vary}.

In the High Mass case, the LMC still produces a faint-dipole signal from the collective response as expected from the dynamical encounter, but the trail of dense, cool material along its orbit is much wider, extending more than 100 kpc along the z-axis, than the fiducial case, and features a steeper gradient in density and temperature along the stream. In addition, more than twice as much LMC CGM gas (but a similar mass fraction) is found in the trailing stream and $\sim 4.5$ times as much within 20 kpc of the LMC disc. The shock driven in the MW CGM remains a robust feature, with a few differences. The CGM of the LMC drives a stronger shock since it retains more gas as it descends into the MW potential well, accelerating more gas to a slightly higher Mach number. The curve of the shock closest to the LMC also appears more round than before, likely a result of the larger surface driving the collision. 

The overall response in the Low Mass case is similar to our fiducial model. However, the lower mass CGM is far less resistant to ram-pressure stripping, so we see that a majority of the LMC's CGM is stripped before crossing $R_{\rm 200, MW}$. This results in less than half of the initial LMC CGM gas trailing along the LMC's orbit within the MW's halo. The densest components near the core do persist long enough to drive a shock, enhancing the density, pressure, and temperature respectively. Since the shock is driven by less material, this also changes the morphology of the shock in the vicinity of the LMC since the supersonic collision with the MW CGM is occurring over a smaller surface area compared to the High Mass and fiducial runs.  

In addition to the Mach number, another aspect of the shock that varies as you change the LMC CGM mass is its stand-off radius, where assuming a larger initial LMC CGM mass produces a larger separation between the shock and the LMC. Following the same procedure in section \ref{subsec:shock}, we locate the stand-off radius at $R_{\rm so} = 0.9$ kpc and $18.2$ kpc for the Low and High Mass runs respectively. The larger stand-off radius from the more massive LMC CGM is primarily from the larger surface area of gas driving the shock (more LMC gas survives infall) and secondly from the modestly higher Mach number of that retained gas. Note that for the real LMC, we would not expect $R_{\rm so}$ to ever be less than what would be the case when only the LMC ISM is included \citep{setton_large_2023}, giving a lower limit on the shock's stand-off radius. 

\begin{figure*}[ht!]
    \centering
  \includegraphics[width=1\linewidth]{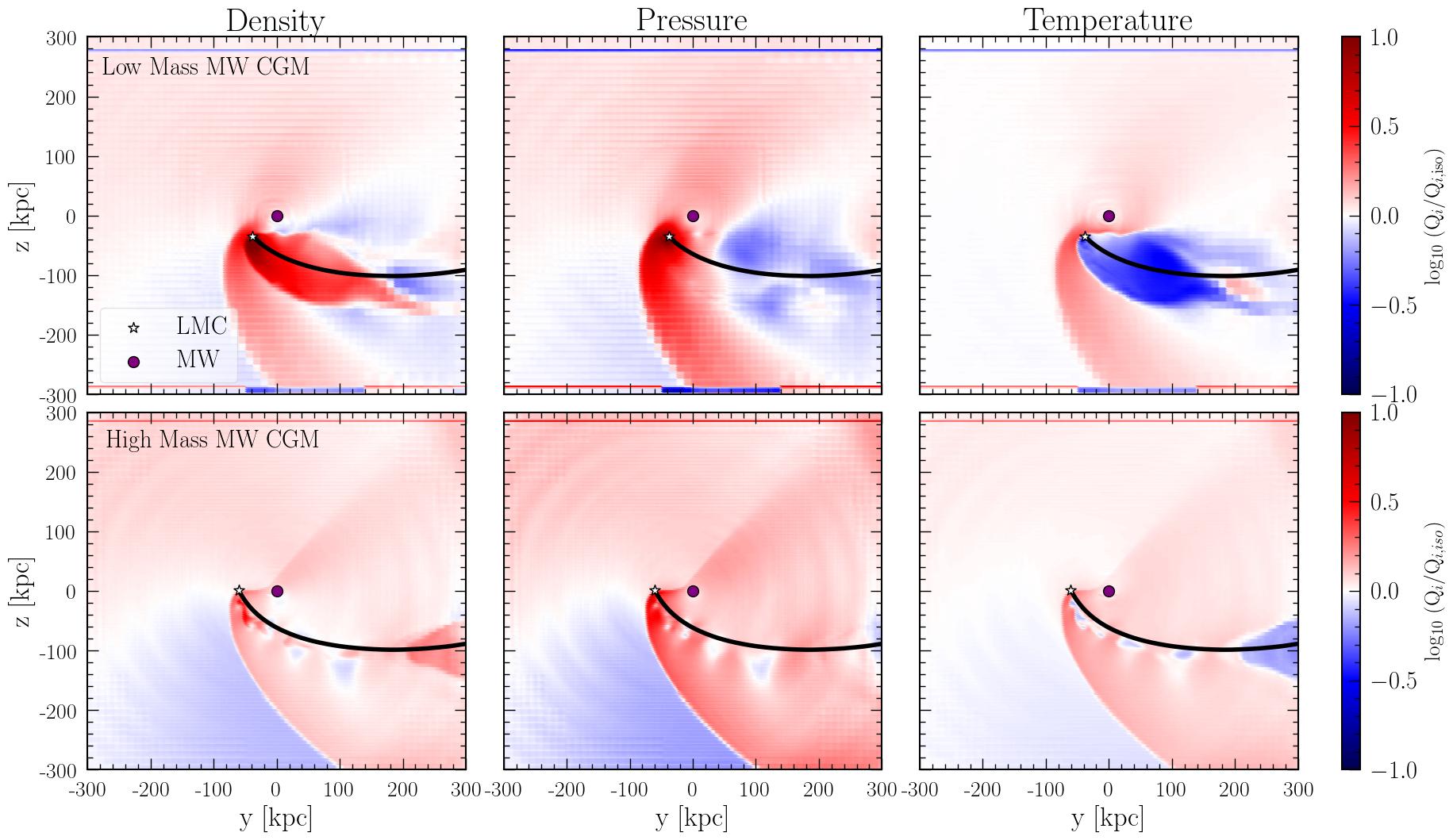}
  \caption{Impact of changing the assumed initial MW CGM mass. Similar to Figures \ref{fig: res} and \ref{fig: resLMC}, but this time, for changes in CGM density (\textbf{left}), pressure (\textbf{center}), and temperature (\textbf{right}) in the Galactocentric y-z plane for both the High Mass (\textbf{bottom}; $M_{\rm CGM,MW} = 6.7 \times 10^{10} M_{\odot}$) and Low Mass (\textbf{top}; $M_{\rm CGM,MW} = 1.2 \times 10^{10} M_{\odot}$) models of the MW CGM. The white star and the black line marks current and past orbital trajectory of the LMC respectively from each respective run to t=3.83 Gyr in simulation time.} 
  \label{fig: resMW}
\end{figure*}

\subsubsection{Varying MW CGM Mass}
The other side of this interaction worth considering is the assumed initial mass of the MW CGM. Following the structure of the prior section, Figure \ref{fig: resMW} presents the changes in the CGM's physical properties for both a High (top) and Low (bottom) Mass MW CGM compared to an isolated MW CGM of the same mass. As summarized in Table \ref{tab: vary}, the MW CGM takes on an initial mass of $M_{\rm CGM, MW} = 6.7 \times 10^{10} M_{\odot}$ and $M_{\rm CGM, MW} = 1.2 \times 10^{10} M_{\odot}$ for the High and Low Mass scenarios respectively. We assume the fiducial value for the LMC CGM mass, yielding almost identical mass ratios from the prior section of $\sim$ 1:28 and $\sim$ 1:5. 

For the High Mass MW CGM comparison, the overall gas response bears a close resemblance to the Low Mass LMC CGM scenario. The higher MW CGM densities enhance the ram-pressure, stripping most of the LMC CGM at early times and leaving less gas to drive the shock. We find less LMC CGM gas in the proximity of the LMC and along the trailing stream compared to the fiducial case but an amount which is comparable to the Low Mass LMC CGM. This leaves $> 70 \%$ of the initial LMC CGM gas in the High Mass MW CGM case at distances greater than 250 kpc from the LMC in the outskirts of the MW halo. The shock shape and amplitude are also similar to the prior scenario with a Low Mass LMC CGM. The overall similarity between the shock properties (shape, stand-off radius) and the efficient stripping of LMC CGM gas points to these signals being useful gauges of the relative masses and characteristic densities of the halo gas surrounding both galaxies. 

The Low Mass MW CGM runs also reflect this point. The LMC produces a comparable shock in amplitude and morphology to the High Mass LMC CGM case, as well as trailing gas structures that mirror each other's wide, extended morphology and steeper (compared to the fiducial case) anti-correlated gradients in density and temperature.

When we compare the stand-off radius, we also find similar distances of $R_{\rm so} = 2.3$ and $24.8$ kpc from the shock to the center of the LMC for the High and Low Mass runs for the MW CGM. In addition to the shape and strength of the shock and the mass/morphology of trailing material, the position of the stand-off radius may serve as an additional diagnostic for the mass ratio between the CGM of the MW and the LMC prior to its infall.  

\newpage
\section{Discussion} \label{sec:disc}

\subsection{Observational Consequences} \label{sec: obs}
Our work shows that the infall of the LMC may be the source of significant enhancements and distortions to the physical and kinematic properties of the MW CGM. These features are marked by their large spatial imprint, provoking a natural response as to whether such all-sky features are observable with existing and near-future probes of the gas surrounding our Galaxy. 

In this section, we consider the possible signals from an enhancement in column density, X-ray surface brightness, and the Sunyaev-Zeldovich effect. To do this, we will show all-sky Mollweide projections of our CGM simulations transformed into Galactic coordinates using \textit{healpy} with NSIDE=64 and a pixel area of $0.84^{\circ}$. This was done for our fiducial simulations with the LMC on its first infall, in addition to our runs with the MW in isolation as a point of comparison. Across all our Mollweide projections, we performed a distance cut, where all foreground gas less than $d_{\rm cut}$ from our observer in the Galaxy is removed. This was done to avoid integration errors that occur when the angular size of nearby gas cells exceeds that of the healpix pixel. We adopt a value of $d_{\rm cut} = 15$ kpc \footnote{The contrast of our all sky signal with the isolated case is not strongly sensitive on the choice of $d_{\rm cut}.$}. We also smooth our maps with a symmetric Gaussian beam with a full-width half-max of $5^{\circ}$ to smooth over any numerical artefacts induced from our AMR grid. 

\begin{figure*}[ht!]
    \centering
  \includegraphics[width=1\linewidth]{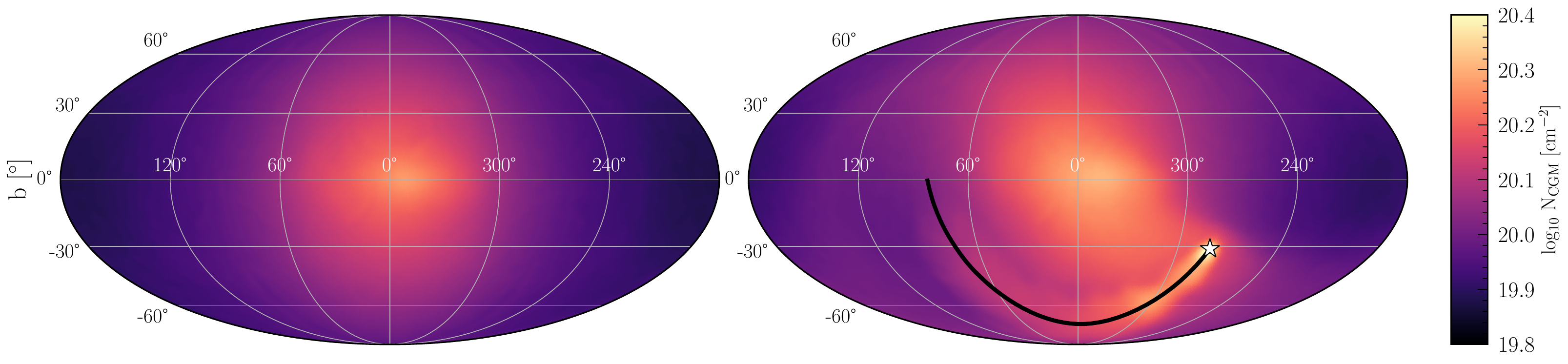}
  \caption{Mollweide all-sky projections of MW CGM column densities in Galactic coordinates for an isolated MW (\textbf{left}) and one with an infalling LMC (\textbf{right}) near the approximate observed position of its stellar disc (\textbf{white star}) and its past orbit (\textbf{black line}). Maps were made using \textit{healpy} with NSIDE=64 and a pixel area of 0.84$^{\circ}$. A Gaussian smoothing with a full-width half-max of 5$^{\circ}$ and a distance cut on all gas $<15$ kpc from the observer at the solar position were applied. The upper and lower bounds on the colorbar match approximately the minimum and maximum of the column densities measured in the MW-LMC map on the right.} 
  \label{fig: logN}
\end{figure*}

\begin{figure}[ht!]
    \centering
  \includegraphics[width=1.0\linewidth]{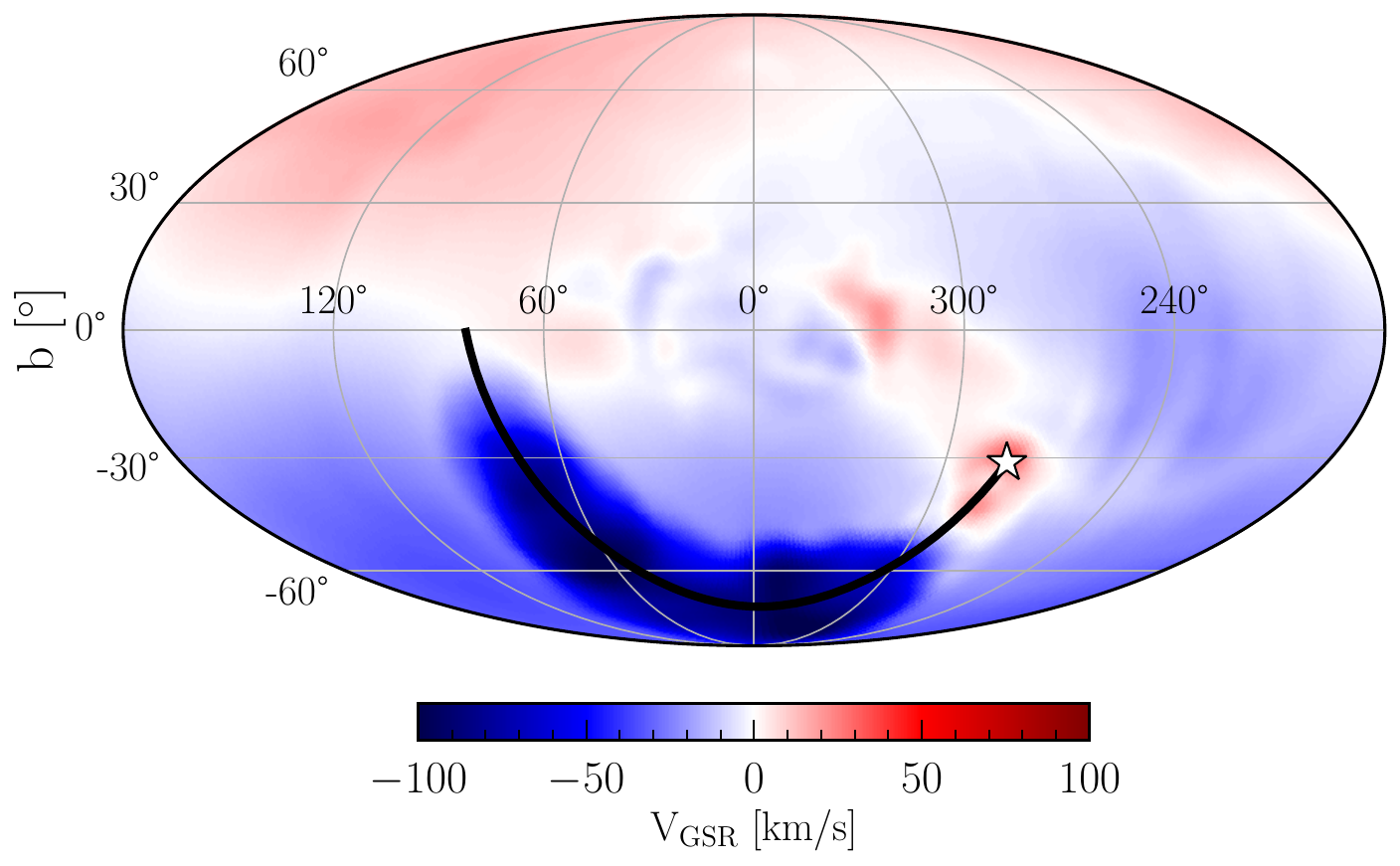}
  \caption{Mollweide all-sky projections of the density-weighted line-of-sight GSR velocities for CGM gas with an infalling LMC near the approximate observed position of its stellar disc (\textbf{white star}) and its past orbit (\textbf{black line}). The reflex motion of the MW in response to the LMC imprints a dipole in velocities across the sky, increasing (decreasing) line-of-sight velocities roughly aligning with the Northern (Southern) hemisphere. Shock-accelerated MW CGM gas and LMC CGM gas still moving with the LMC are marked by positive velocities while stripped gas from the LMC CGM traces the past orbit of the LMC with large negative velocities $\lesssim$ $-100$ km/s. The contribution from the sun's motion has been subtracted.} 
  \label{fig: vlos}
\end{figure}

\subsubsection{Column Densities}
First, we will consider the spatial fluctuations in matter along all lines of sight through our simulation. In Figure \ref{fig: logN}, we show all-sky maps of the column density $N_{\rm CGM}$ through the CGM for the MW in isolation and with the infalling LMC, with its stellar disc marked by the white star in Galactic coordinates at the end of its orbital trajectory up to our present-day snapshot. In the isolated case, we see little to no features in the column density beyond the increase in density at the Galactic Center from the overall density profile. When we include the LMC, the shock and the stripped material enhance column densities across the Southern hemisphere. 

The shock appears as a sharp discontinuity in the column density, offset from the center of the LMC at ($l,b$)=(280.56, -31.16), and arching from the LMC to the vicinity of the Galactic center. We can get a rough estimate of the jump by comparing the measured column densities through sightlines behind and ahead of the LMC. Using the Galactic $b$ latitude of the LMC, we find a column density contrast of $\sim 0.2$ dex across the shock for sightlines $\pm 30^{\circ}$ in Galactic longitude $l$. In addition to the shock, the stripped material of the LMC CGM reaches column densities of log$_{10} N_{\rm CGM} \gtrsim 20.3$ cm$^{-2}$ immediately trailing the LMC and declines by $\sim0.3$ dex along the past orbit of the LMC. In our all-sky projections, this material is largely confined to $b < -20^{\circ}$ and  $60^{\circ} \gtrsim l \lesssim 120^{\circ}$. 

Figure \ref{fig: vlos} shows the density-weighted line-of-sight velocities in the Galactic Standard of Rest (GSR) frame for MW CGM gas with an infalling LMC. This was done by subtracting off the GSR velocities for an isolated MW CGM from our MW-LMC runs, removing any contribution from the sun's motion and leaving only the velocity structure imprinted from the MW-LMC encounter. Just as we explored in section \ref{sec: vels}, the kinematic response in the MW CGM manifests as a dipole in radial velocities, increasing (decreasing) velocities in the Northern (Southern) hemispheres by $\Delta$V$_{\rm GSR }$ $\sim$ 20-30 km/s from the reflex motion of the MW. The spatial extent of the reflex motion signal in Figure \ref{fig: vlos} bares a close resemblance to its reported detection in GSR velocities in the stellar halo \citep[see Figure 3 in][]{Chandra2024}. We note that the all-sky velocity signal in the CGM is particularly sensitive to the choice of $d_{\rm cut}$, as the reflex motion signal grows larger for gas at greater distances from the MW. 

In addition to this gravitational response, we also see in Figure \ref{fig: vlos} that the LMC shock accelerates MW gas to positive velocities similar to LMC CGM gas in close proximity to its stellar disc, but the most prominent kinematic signature is the stripped gas of the LMC CGM along the past orbit of the LMC, moving at V$_{\rm GSR} \lesssim -100$ km/s. Both the velocities and spatial extent of the stripped LMC CGM gas is consistent with the high-velocity UV absorbing gas detected in the Southern hemisphere potentially associated with the extended MC Stream \citep{kim_identifying_2024}. This highly ionized gas has been detected primarily through OVI absorption (indicative of diffuse gas at $T\sim10^{5.5}$ K) with \textit{FUSE} and later with HST-COS sightlines \citep{sembach_highly_2003,richter_hstcos_2017}. Although this region of sky also overlaps with ionized gas in the Local Group believed to be affiliated with the M31 system \citep{Lehner2020}, our work suggests that a considerable fraction of this high-velocity ionized material in the Southern hemisphere may have originated in the CGM of the LMC and is potentially at great distances, $\sim 100$ kpc from the MW. 

This fast-moving LMC CGM material, and possibly components of the shock, may not only be detectable in absorption, but also in UV emission \citep[e.g.][]{Corlies2016}. Existing space-based missions probing UV emission lines, specifically OVI and CIV lines from $T\sim 10^{5-6}$ K gas across large regions of the sky, such as FIMS/SPEAR \citep{Jo2019}, and future SmallSat missions like Aspera \citep{Chung2021}, will enable future work comparing model predictions for the total UV emissivity of MW and LMC CGM gas to these data. Observing the stripped LMC CGM gas in both absorption and emission will also better constrain its distance, and if such material traces the LMC's past orbit, it could open the door to using CGM gas in the MW halo as an indirect probe of the Clouds' dynamics and the Galactic potential. 

Another probe of inhomogeneities in the MW CGM along different lines of sight that may hold promise are fast radio bursts (FRBs) \citep{prochaska_probing_2019,Platts2020,Cook2023}. Although the physical origins of FRBs are not well understood, the interaction between free electrons and photons in the intervening plasma between the FRB and the observer causes a delay in the arrival times of the initial short burst of radiation as a function of frequency. Thus a precise measurement of the dispersion measure of photon arrival times as a function of frequency encodes information on the integrated electron number density along a given line of sight. Existing surveys such as the Canadian Hydrogen Intensity Mapping Experiment \citep[CHIME,][]{CHIME2018}, the Deep Synoptic Array-110 (DSA-110) and its successor the DSA-2000 \citep{Hallinan2019} are expected to dramatically expand the number of localized FRBs in the Milky Way and beyond, offering a novel probe of CGM anisotropies potentially induced by the recent collision of massive satellites like the LMC.

\begin{figure*}[ht]
    \centering
  \includegraphics[width=1\linewidth]{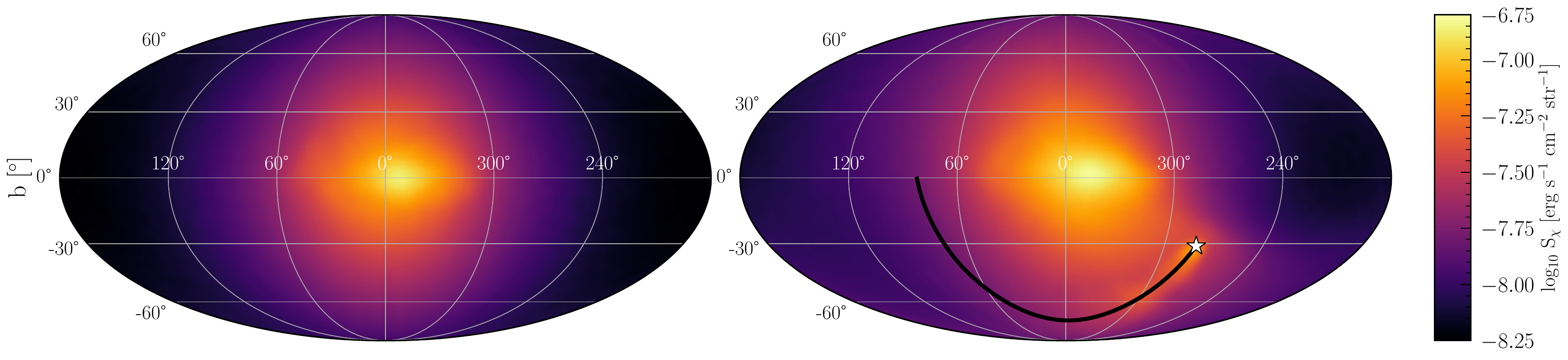}
  \caption{Similar to Figure \ref{fig: logN}, but now showing Mollweide all-sky projections of the X-ray surface brightness in the bandwidth range of (0.2 - 7) keV in Galactic coordinates for an isolated CGM (\textbf{left}) and one with an infalling LMC (\textbf{right}) near the approximate observed position of its stellar disc (\textbf{white star}) and its past orbit (\textbf{black line}). All gas is assumed to have a uniform metallicity of $Z_{\rm CGM} = 0.3 Z_{\odot}$. The upper and lower bounds on the colorbar match approximately the minimum and maximum of the X-ray surface brightness measured in the MW-LMC map on the right.} 
  \label{fig: Xray}
\end{figure*}

\subsubsection{X-Rays}
The supersonic collision between the MW and LMC produces a shock that both compresses and heats the $T \sim 10^6$ K ambient MW CGM gas. This increase in density and temperature across the shock may enhance the X-ray luminosity of this gas. In Figure \ref{fig: Xray}, we show all-sky maps in Galactic coordinates of the X-ray surface brightness of the CGM in isolation and with the infalling LMC. We estimate the surface brightness by integrating over the X-ray emission of each gas cell across all lines of sight:
\begin{equation}
    \begin{aligned}
    S_{\chi} = \int n^2 \epsilon_{\chi}(T,Z) dl \: [\text{erg} \text{ s}^{-1} \text{ cm}^{-2} \text{ str}^{-1}]
    \end{aligned}
\end{equation}
We use the X-ray emissivity, $\epsilon_{\chi}$(T,Z), from the atomic spectroscopy analysis code \textit{Chianti} \citep{Chianti2020}, where we assume a uniform metallicity of $Z_{\rm CGM} = 0.3 Z_{\odot}$ and a bandwidth of (0.2 - 7) keV. The integral has units of erg s$^{-1}$ cm$^{-2}$, and we divide by $4\pi$ steradians to arrive at an X-ray surface brightness across the sky. 

We find that the LMC's shock front brightens the X-ray sky, boosting the surface brightness across an extended region on the LMC's side of the Southern hemisphere. The contrast between the same sightlines used in the prior section, $\pm 30^{\circ}$ in galactic longitude from the coordinate position of the LMC, reaches $\sim 0.4$ dex in brightness estimates measured behind and ahead of the shock. The compressed, stripped CGM material of the LMC also brightens in X-rays, however uncertainties surrounding its metallicity and temperature will affect its predicted brightness.

All-sky maps of the X-ray surface brightness have been completed by space-based telescopes such as ROSAT \citep{Snowden1997} and recently eROSITA \citep[eRASS,][]{Zheng2024_erass, Zheng2024_erass1}, in addition to X-ray studies of the diffuse emission surrounding the LMC \citep{gulick_total_2021,Locatelli2024}. For a surface brightness of $S_{\chi} \sim 10^{-7}$ erg s$^{-1}$ cm$^{-2}$ str$^{-1}$ and an assumed average photon energy of $0.2$ keV, this leads to an expected photon flux of order unity per second per square degree. However, any signal from the shock would be competing from other extended bright X-ray signals in the Southern hemisphere, such as the Fermi-eROSITA bubbles \citep{Predehl2020}, which extend down to $b$ $\sim$ -60$^{\circ}$. 

\begin{figure*}[ht!]
    \centering
  \includegraphics[width=1\linewidth]{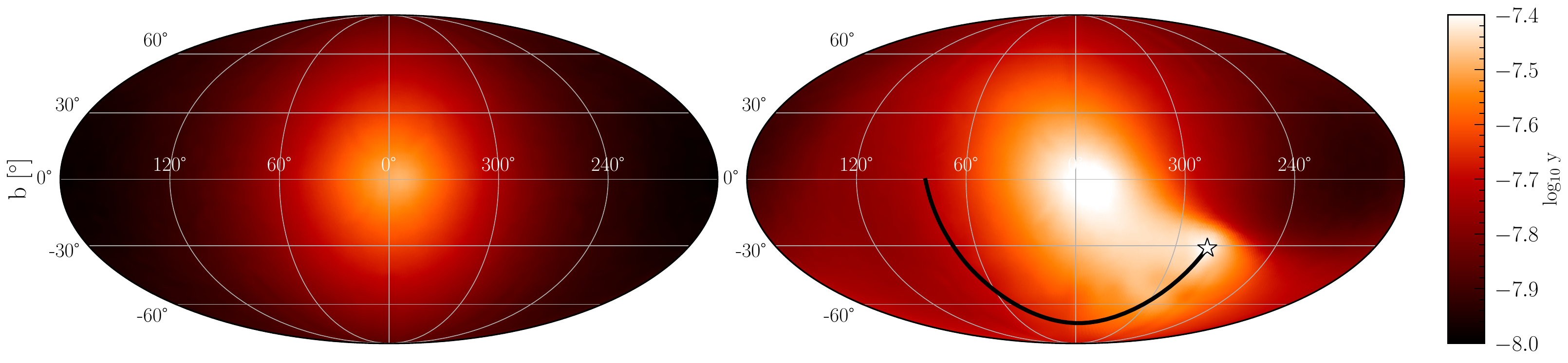}
  \caption{Similar to Figures \ref{fig: logN} and \ref{fig: Xray}, but now showing Mollweide all-sky projections of the tSZ Compton-y parameter of the MW CGM in Galactic coordinates for an isolated MW (\textbf{left}) and one with an infalling LMC (\textbf{right}) near the approximate observed position of its stellar disc (\textbf{white star}) and its past orbit (\textbf{black line}). The upper and lower bounds on the colorbar match approximately the minimum and maximum of the Compton-y measured in the MW-LMC map on the right.} 
  \label{fig: tSZ}
\end{figure*}

\subsubsection{The Sunyaev-Zeldovich (tSZ) Effect}
The last potential observable of the large-scale response of the MW CGM to LMC infall we will consider in this work is the spectral distortion of the Cosmic Microwave Background (CMB) from scattering events between CMB photons and fast-moving electrons in the hot halo, a phenomena known as the (thermal) Sunyaev-Zeldovich (tSZ) Effect \citep{SZ1969}. The strength of this spectral distortion from the tSZ, quantified in terms of the Compton-$y$ parameter, can be used as a direct measure of the thermal pressure of the gas along a given line of sight:
\begin{equation}
    \begin{aligned}
        y = \frac{\sigma_T}{m_e c^2} \int P_e dl,
    \end{aligned}
\end{equation}
where $\sigma_T$ is the Thomson cross section, $m_e$ is the mass of an electron, and $c$ is the speed of light. The tSZ effect has proven to be a robust measure of shocks and other pressure variations in the ICM \citep{Mroczkowski2019}. If the largest discontinuity across the shock is in the gas pressure, then the shock produced from the LMC should generate some spatial-varying enhancements to the all-sky contribution to the tSZ from the MW CGM. 

In Figure \ref{fig: tSZ}, we plot the estimated all-sky Compton-y signal from our MW CGM simulations using the integrated gas pressure along every line of sight. As expected for the isolated case, the MW hot halo contribution is very small, with characteristic values of $y\sim10^{-8}$ except near the dense core, in line with previous analytical estimates of the local MW CGM contribution of $y \sim 5 \times 10^{-9}$ \citep{Khatri2015}. However, when we include the infall of the LMC, the collision leads to a sharp excess in the tSZ signal along the curvature of the shock, with Compton-y values approaching $\sim10^{-7.4}$ across a sizeable portion of the Southern sky. 

We leave a more detailed calculation of the detectability of the tSZ signal from the shock to future work. We anticipate that the signature in tSZ from the hot CGM will be quite small and difficult to detect directly, particularly for ground-based observatories, due to its large angular scale. However, it may be possible to detect cross-correlations in tSZ with existing all-sky Compton-y maps \citep{Planck2016,ACT2020,Tanimura2022,SPT2022,McCarthy2024,ACT2024}, as well as make predictions for upcoming ground-based experiments from the Simons Observatory \citep{Abe2019} and CMB-S4 \citep{CMB-S42016} and future space-based telescopes such as PIXIE \citep{PIXIE2024} and LiteBIRD \citep{LiteBIRD2024}. This prediction of the tSZ enhancement from the LMC's shock front affirms the capability of the tSZ signal as a potential probe of the pressure substructure in our own hot halo, and future CMB satellite experiments with large-scale sensitivity may be best suited to detect the tSZ signal from the shock. 

\begin{figure}[h!]
    \centering
  \includegraphics[width=1\linewidth]{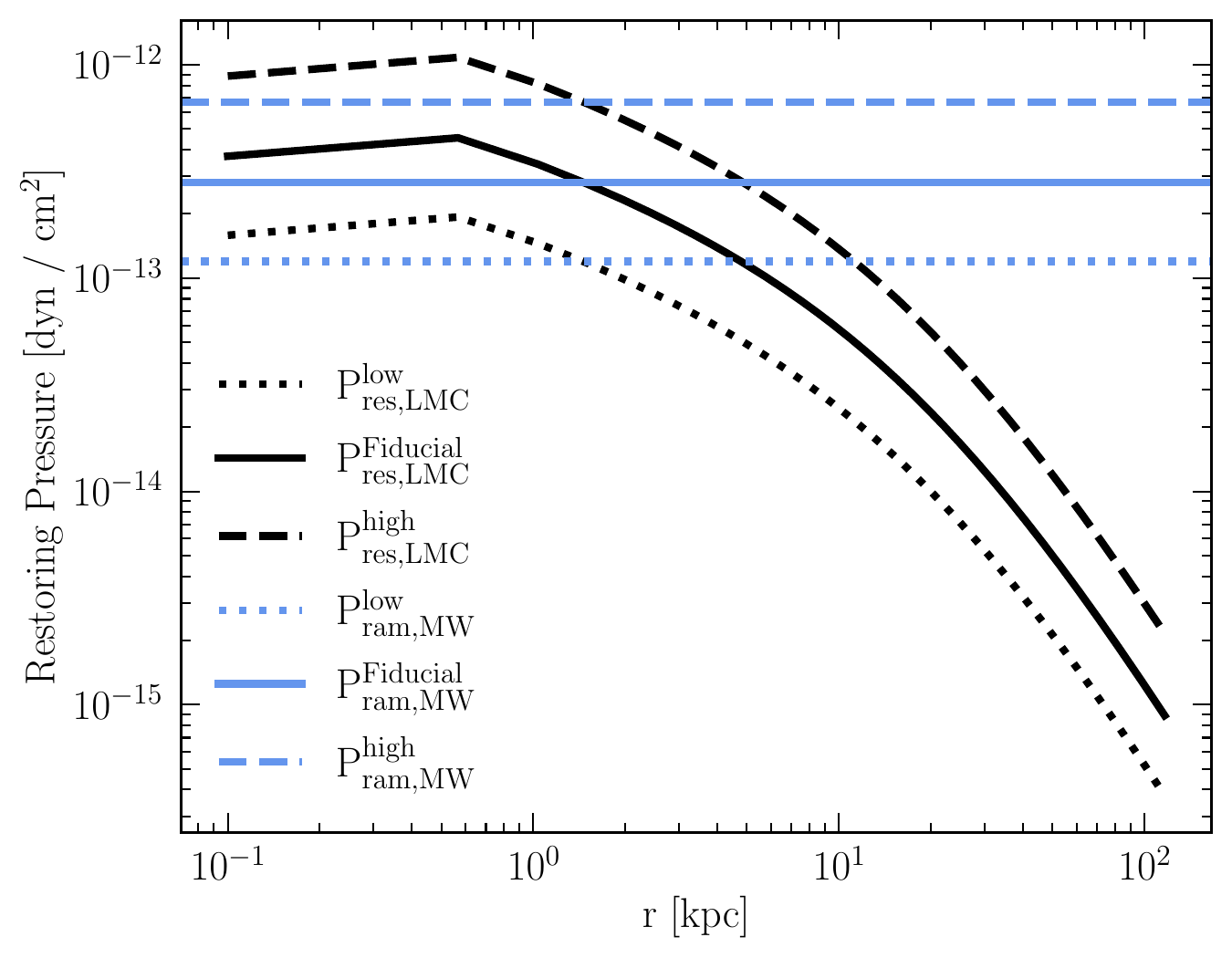}
  \caption{Comparison to the criteria for instantaneous ram-pressure from \cite{McCarthy2008}. Black lines trace the gravitational restoring pressure as a function of projected radius of each of the three LMC CGM density profiles explored in this work: the Fiducial ($M_{\rm CGM,LMC}=2.4 \times 10^{9} M_{\odot}$), the Low Mass ($M_{\rm CGM,LMC}=1 \times 10^{9} M_{\odot}$), and the High Mass runs ($M_{\rm CGM,LMC}=5.7 \times 10^{9} M_{\odot}$). The blue horizontal lines mark the upper threshold for instantaneous ram-pressure stripping from each MW CGM run: the Fiducial ($M_{\rm CGM,MW}=2.8 \times 10^{10} M_{\odot}$), the Low Mass ($M_{\rm CGM,MW}=1.2 \times 10^{10} M_{\odot}$), and the High Mass runs ($M_{\rm CGM,MW}=6.7 \times 10^{10} M_{\odot}$). Where the lines intersect defines the stripping radius, R$_{\rm strip}$.}
  \label{fig: strip}
\end{figure}

\subsection{Survival of the LMC's CGM} \label{sec: strip}
One adjacent question to our work is how long do we expect the gaseous halo of the LMC to survive under ram-pressure. The CGM of infalling dwarf galaxies into the CGM or ICM of more massive host systems are not expected to survive long in these environments due to their weak restoring force from gravity \citep{zhu_its_2024}. However, for more massive satellites like the LMC, the simulations of \cite{lucchini_magellanic_2020,lucchini_magellanic_2021,lucchini_properties_2023} and subsequent observations from \cite{krishnarao_observations_2022} indicate that the CGM of the LMC may source the $\sim 10^9 M_{\odot}$ reservoir of ionized gas surrounding the Clouds and the Stream. 

Here we conduct a similar test to that done in \cite{zhu_its_2024} where we compare our model to the instantaneous analytical ram-pressure stripping criteria for gaseous halos introduced in \cite{McCarthy2008}. The criteria relates the ram-pressure $P_{\rm ram} = \rho_{\rm CGM} v_{\rm orb}^2$ at a given time to the gravitational restoring force per unit area of the infalling galaxy from its total mass and gas profiles as a function of its projected distance $R$ from its center: 
\begin{equation}
    \begin{aligned}
        P_{\rm ram} (t) > \alpha \frac{G M_{\rm tot} (R) \rho_{\rm gas} (R)}{R},
    \end{aligned}
\end{equation}
where $\alpha$ is a dimensionless factor of order unity describing the exponential falloff of the gas profile $\alpha \approx 3 \beta$. The radius where the two sides of this inequality meet defines the stripping radius $R_{\rm strip}$, where at smaller radii the gravitational restoring force of the infalling galaxy is able to resist ram-pressure stripping. 

In Figure \ref{fig: strip}, we apply this ram-pressure stripping criteria to the LMC at pericentre for the different realizations explored in the previous section. For our fiducial MW model where we vary the LMC CGM mass, we find $R_{\rm strip}$ = 1.5, 0, and 4.8 kpc for the Fiducial, Low Mass and High Mass runs respectively. For our runs where we vary the MW CGM mass and keep the LMC CGM mass fixed, we find the same values of $R_{\rm strip}$: 1.5 kpc for the Fiducial run, and $4.8$ kpc and 0 kpc for the Low Mass and High Mass MW CGM. Across this limited exploration of the simulated parameter space of MW/LMC CGM masses, $R_{\rm strip}$ are all smaller than the LMC HI disk measured at $\sim 6$ kpc \citep{salem_ram_2015} and correlated with the initial MW/LMC CGM mass ratio. 

This suggests that the vast majority of the LMC CGM contribution to the ionized MS should be unbound from the LMC, in either close proximity to the disc or confined to its trailing side, in agreement with the mass fractions listed in Table \ref{tab: vary}. Since we do not include an ISM component, we cannot capture the ram-pressure directly applied to its HI disk, but if its CGM has only recently been cleared on its leading edge in the last several $\sim$Myr as \cite{zhu_its_2024} and our work suggests, then the direct interaction between the LMC and the MW CGM and the observed truncation of the LMC's HI disk relative to its stellar disk are a recent developments. The compressive front of the leading edge from ram-pressure may ignite renewed star formation episodes in the LMC \citep{de_boer_bow-shock_1997,Harris2009}. This means that the star formation history of the LMC could serve as an indirect probe of the survival time of its gaseous halo. 

This, of course, is all under the assumption that the LMC has recently completed its first pericentre infall. A previous passage scenario would mean that all or most of the LMC CGM should have been removed on that previous infall and there shouldn't be as much of an ionized gas tail in the current stream on this more recent passage\citep{vasiliev_dear_2023}. In addition, its HI disk would have been directly exposed to the ram pressure of the MW CGM for a much longer period of time. 

\subsection{Comparison to Previous Work} \label{sec: sims}
Now we consider our work in the context of existing simulations, beginning with those that model aspects of the MW-LMC system. First are the wind-tunnel simulations introduced in \cite{salem_ram_2015} but later used in \cite{setton_large_2023} to study the bow shock caused from the supersonic collision of the LMC HI disk and the MW CGM. They find an asymmetric $\mathcal{M} \approx 2$ shock approximately $\sim 30$ kpc in length that produces jumps in density and temperature consistent with the RH jump conditions. Their LMC simulations are initialized with a very low-mass CGM, $\sim 10^6 M_{\odot}$, that is swept away early in their runs, leaving only the HI disk to drive their shock. However, our work shows that the inclusion of the LMC CGM, with a total mass comparable to the LMC stellar mass, drives a much larger shock front, on the scale of $\sim R_{200, MW}$, that is shaping a large portion of the MW CGM's gas. Future work with LMC models that include both a massive ISM and CGM component will allow us to better disentangle these overlapping shocks. MW CGM models with multiphase substructure will also be required to assess the shock's ability to ionize neutral substructure in the MW CGM and produce the extended $H\alpha$ emission and observed line ratios associated with the Clouds and the Stream \citep{wakker_characterizing_2012}. 

In \cite{lucchini_properties_2023}, they outline the details of their LMC CGM models adopted in their earlier studies. Their LMC CGM is modeled with an initial mass of $> 5 \times 10^{9} M_{\odot}$, about $1/4$th the total mass of their MW CGM. The authors are able to reproduce many aspects of the Stream, such as its massive reservoir of ionized gas, its atomic/ionized morphology, and its ionization fractions. Although there are many modeling aspects included in their work that we do not currently have, such as the SMC, radiative cooling, star formation, etc, our work does support the idea introduced by \cite{lucchini_magellanic_2020} that a fraction of the ionized gas trailing the LMC along the past orbit could have originated in its CGM. 
Roughly $\sim 10^9 M_{\odot}$ of ionized gas is observed in the trailing stream associated with the MC Stream at an assumed distance of 55 kpc \citep{fox_cosuves_2014}, but the total mass increases to $\sim 5 \times 10^{9} M_{\odot}$ if the gas is assumed at a common distance of 100 kpc. This is broadly consistent with our High Mass LMC CGM model under the assumption all of this ionized gas originated in the LMC CGM. However, the CGM properties that we should expect for LMC-like galaxies, let alone those of the LMC prior to infall, are highly unconstrained and different assumptions on the nature of SNe feedback give a range of predictions of how much CGM gas these low-mass galaxies should actually possess \citep{Christensen2016,hafen_origins_2019,Carr_2023ApJ...949...21C, CrainVdVoort2023}. More work is needed to distinguish between the LMC CGM and other possible contributions to the ionized budget from MW CGM mixing and the shock. 

Another difference between \cite{lucchini_properties_2023} and the work presented here is how the LMC CGM is distributed about the LMC. Estimates of $R_{\rm strip}$ from \cite{zhu_its_2024} for the LMC CGM models of \cite{lucchini_magellanic_2021} find stripping radii larger than the LMC HI disk, meaning that LMC CGM gas should be both trailing and leading the LMC past its pericentre. The survival of the LMC CGM on its leading side may be a product of their assumed slope of the MW CGM density profile ($\beta \sim 0.55$), which is slightly steeper than we assume in this work, resulting in less ram pressure stripping as the LMC descends into the halo. However, the truncation of the LMC's HI disk relative to its stellar disk suggests that the ISM is under direct ram pressure from the MW CGM \citep{salem_ram_2015}, implying that the lower density LMC CGM gas would already have been cleared on its leading side. This view is also supported by the reported detection of the LMC CGM in UV absorption through sightlines all near the LMC or confined to its trailing side \citep{krishnarao_observations_2022}. More sightlines probing the leading side of the LMC would aid in constraining the spatial distribution of any remaining LMC CGM gas, as well as the location of the shock.

Stepping away from the MW-LMC system, ram-pressure stripping of satellite gas and induced cooling have been invoked as a possible seeding mechanism for $T\sim 10^4$ K cold gas at large radii ($r > 0.5 R_{\rm 200}$) seen in CGM simulations of MW-like galaxies in cosmological simulations but not in isolated ones \citep{Fielding2020, roy_seeding_2023}. Although these works did not include satellites as massive as the LMC with their own CGM, the large spatial extent of the LMC's stripped CGM gas from beyond $\sim R_{\rm 200, MW}$ to its present position is consistent with this picture of satellite-seeding of cool gas and subsequent cooling in the mixing layers with the ambient MW CGM to form a cold gas reservoir \citep{tonnesen_its_2021}. Satellites can also induce local thermal instabilities in the hot phase of the CGM from strong density perturbations, leading to the condensation of cold gas out of the volume-filling hot phase \citep{Field1965,mccourt_thermal_2012, sharma_thermal_2012,singh_cold_2015,saeedzadeh_cool_2023}. The criteria for this condensation and its dependence of the ratio of the local cooling time and freefall time of the gas is still not well understood \citep{palchoudhury_multiphase_2019}, but in the FIRE-2 simulations for example, local perturbations that enhance the density by greater than unity (and by consequence, lowering its cooling time) have been found to source most of the cool phase in the CGM and fuel the central galaxy with star-forming gas \citep{hafen_fates_2020,esmerian_thermal_2021}. The strongest density enhancements in the CGM from our work are found in the stripped LMC CGM gas and in the shock front, and future work that includes radiative cooling will show how these features of the interaction affect the multiphase structure of the MW CGM and populate its cooler gas reservoirs. 

The first infall of the LMC is a chance to study merger-induced shocks in the CGM around MW-like galaxies. Such shocks are typically only studied in the context of the ICM or the CGM of massive ellipticals, due to the brighter X-ray luminosity and stronger tSZ signal from high $\mathcal{M}$ shocks \citep[for review see][]{Markevitch2007}. Merger-induced shocks from massive satellites have the potential to heat the gas of the CGM and excite turbulence like that observed in the CGM of low-z galaxies \citep{Werk2013,Chen2023} and in zoom-in cosmological simulations of similar systems \citep{Lochhaas2022}.

\begin{figure*}[ht!]
    \centering
  \includegraphics[width=1\linewidth]{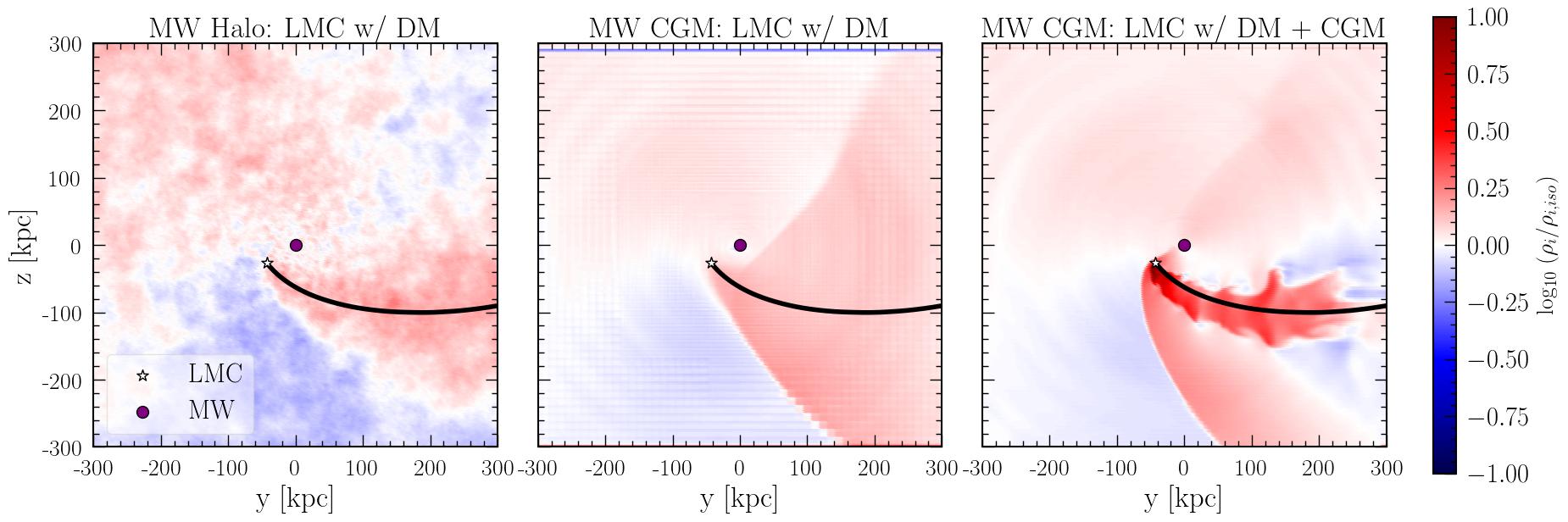}
  \caption{Changes in density for the MW DM halo (\textbf{left}), the MW CGM with a DM-only LMC (\textbf{center}), and including the LMC CGM (\textbf{right}; identical to the density map from Figure \ref{fig: res}). The collective response of the halo is present in both the DM and the CGM, and the dynamical friction wake is prominent in the MW DM halo and in the CGM with a DM-only LMC. Stripped LMC CGM gas overlaps with the density wake and enhances the density in the shock.}  
  \label{fig: resDM}
\end{figure*}

\subsection{Implications for the Milky Way System} \label{sec: Mwres}
In this section, we reflect on the broader implications the infall of the LMC and the MW CGM response may have on other aspects of the MW system. This includes the DM and stellar components of the Galactic halo, the gaseous properties of the inner MW CGM/ISM, as well as the satellite populations of the MW and LMC.  
\subsubsection{The Milky Way Dark Matter \& Stellar Halo}
Our work shows that the CGM, the collisional component of the halo, is deeply impacted by the large and recent encounter between our Galaxy and the LMC, a response that is reflected in the collisionless dark matter halo. In Figure \ref{fig: resDM}, we show the overdensities in the MW DM halo on the left, compared to the CGM overdensities with (right) and without an LMC CGM (center). The overdensities in the dark matter are measured with respect to the MW in isolation, similar to the method described for the gaseous properties in section \ref{subsec:res}. The MW DM response to LMC infall matches the signal measured in \cite{garavito-camargo_hunting_2019}, producing both transient and large-scale collective wakes. When we do not include the LMC CGM, we find a comparable response in the CGM density to that seen in the DM from both the collective response and a gaseous dynamical friction wake \citep{Ostriker1999}. It isn't until we include the LMC CGM that we see the overlapping contributions from both the large-scale dynamical response and the contributions from the shock and the stripped gas along the past LMC orbit, enhancing the overdensities predicted from gravitational density wakes alone.  

The reflex motion signal, which takes the form of redshifted and blueshifted radial velocities in the Northern/Southern hemispheres, has reportedly been detected in measurements of the stellar halo, and our work shows that such a signal may also be present in the velocity measurements of the MW CGM. Most of the cloud structures where we have good distance measurements are $\lesssim 10$ kpc away \citep{Lehner2022}, and thus the reflex motion is not expected to contribute much to their observed velocities. However, for more distant structures, exceeding $r \gtrsim 50$ kpc, the reflex motion may be a considerable contribution to the measured radial velocities of gas in both hemispheres. 

\subsubsection{The Milky Way's Inner Circumgalactic Medium and Interstellar Medium}
The large surface area of the LMC's shock front means that regions of the MW CGM can still be influenced by the LMC's passage, despite not falling directly on its past orbit. We see an example of this with the northern segment of the LMC's mach cone, which shares a cross-section with the inner MW CGM and potentially the MW ISM. Despite the downstream contrast with the wave front being weaker than what is measured in the Southern Hemisphere, it still raises an interesting question as to whether these large-scale shock/wave fronts interact with other structures in the inner Galaxy. Since our model does not include an ISM component or the active contribution from star formation \& AGN feedback, we do not accurately capture the complex multiphase environment of the inner MW CGM. However, the shock wave of the LMC may be an additional contributor, shaping the morphology of nearby cold clouds, driving turbulence \& other kinematics, and perhaps other emerging large-scale structures, such as the Fermi-eROSITA Bubble \citep{Su2010,Predehl2020}. 

\subsubsection{Satellites of the Milky Way and LMC}
The global CGM response from the LMC means that not only is the MW affected but also potentially the properties of other satellites of the MW and the LMC. Using orbital histories calculated from \cite{patel_orbital_2020}, \cite{setton_large_2023} determined that two nearby ultrafaint dwarfs (Car3 and Hyil) are likely to have spent $\sim500$ Myr in the proximity of the shock and that the dwarf, Ret2, may be inside the shock in the present day. Due to the density enhancement by a factor of 2 across the boundary, this could lead to an enhancement of the ram-pressure experienced by these low-mass galaxies, either hastening the stripping of interstellar gas \citep{tonnesen_impact_2008} or perhaps even a moderate enhancement of star formation \citep{zhu_when_2024}. Given the larger shock front predicted from a CGM-CGM collision, we suspect that this effect may extend not only to satellites associated with the LMC, but potentially to other satellites of the MW across the Southern Hemisphere. One possible consequence is whether the enhancement of ram-pressure across the shock influences satellite quenching fractions not only as a function of Galactocentric radius but also of a function of azimuthal angle. This could be a general prediction not only for our Galaxy but for other massive galaxies hosting LMC-like satellites.  

Although not included in this work, the SMC, the LMC's less massive companion, is also expected to have spent considerable time near the shock front and the LMC's CGM. A rich avenue of future inquiry could be to investigate how the dynamic history of the SMC through this inhomogeneous environment surrounding the LMC could be imprinted in the temporal and spatial features of its star formation history \citep{Rubele2018,Sakowska2024}. In addition, the measured speed of the SMC ($\sim 250$ km/s) suggests that it too should form a weak bow shock from its supersonic collision with either the MW or even LMC CGM, which could affect the shape and amplitude of the LMC's shock front in regions where the two features overlap or the morphology of the LMC's stripped material. 

\subsection{LMC Shock Acceleration of Cosmic Rays} \label{sec: CRs}
The merger-induced shock from the MW-LMC CGM collision may also shape the energetics of high-energy particles in the MW CGM. Shocks are active sites of acceleration for relativistic electrons and protons in a process known as diffusive shock acceleration (DSA) \citep[e.g.][]{Bell1978a,Bell1978b}. Through DSA, scattering events between non-thermal particles and magnetic fields accelerate cosmic rays confined in the shock as they make several crossings between the upstream and downstream flows. The acceleration of high-energy electrons produces synchrotron emission from shocks, which has been observed in both supernova remnants \citep[e.g.][]{Berezhko2004} and in the radio relics of cosmological shocks in the ICM \citep[e.g.][]{Hoeft2007}, in addition to gamma-ray emission from accelerated proton collisions \citep{Ackermann2013}. 

This results in a power-law distribution of cosmic ray energies, $n_E \propto E^{-s}$. The spectral index, $s$, of the accelerated population is related to the compression ratio of the shock, $s = (r+2)/(r-1)$, where $r \equiv \rho_1 / \rho_0$. For the shock produced from the LMC CGM, we find a compression ratio of $r\sim2$ in our fiducial run, yielding a steep spectral index of $s=4$. This falloff suggests there will be few injected high-energy particles that will emit synchrotron radiation, however the expected radiation could be boosted if the shock re-accelerates a pre-shock population of fossil cosmic rays \citep[e.g.,][]{Kang2020}. The estimated synchrotron emission from the shock and the associated brightness temperature are also strongly dependent on the magnetic field structure of the MW CGM, the fraction of thermal energy that goes into accelerating cosmic rays, and the effective observing beam size, etc \citep[e.g.][]{Hoeft2007}. More detailed work is needed to estimate the radio emission from the LMC shock, but its large physical size suggests that it could be an important contributor to the extended synchrotron emission from the MW CGM over large angular scales in the Southern hemisphere. 

\section{Summary \& Conclusion} \label{sec:summ}
In this study we illustrate how the first passage of a massive satellite like the LMC can leave its mark on the galactic atmosphere of its host galaxy. We ran a suite of simulations of a MW-like CGM embedded in a live NFW DM halo with an infalling LMC-like satellite possessing its own CGM to understand how the LMC is shaping the global physical and kinematic state of the MW CGM. We compared the properties of the MW CGM with and without an infalling LMC to isolate the contributions from the encounter, and characterized the shock front produced from the supersonic collision. The major conclusions of our work are listed below: 
\begin{itemize}
    \item The infall of the LMC with its own CGM sources order-unity distortions to the MW CGM in density, temperature, pressure from a combination of gravitational and collisional hydrodynamic effects. This includes (i.) sharp discontinuities in density, temperature, pressure, and entropy across the shock from the supersonic collision between the MW and LMC CGM, (ii.) $\sim 10^{8-9} M_{\odot}$ of warm, ionized gas along the past orbital trajectory of the LMC stripped from its CGM, and (iii.) a dipole-like collective response from the reflex motion of the MW from the outer halo and towards the past pericentre position of the LMC. The following three bullet points expand on these features.
    \item The jumps in density, pressure, temperature, and entropy across the shock are consistent with a $\mathcal{M} \approx 2$ shock. In our fiducial model that assumes $M_{\rm CGM, MW} = 2.8 \times 10^{10} M_{\odot}$ and $M_{\rm CGM, LMC} = 2.4 \times 10^9 M_{\odot}$ at t=0, the size of the shock front exceeds $\sim R_{\rm 200, MW}$ with an estimated stand-off radius of 7.4 kpc from the LMC center, which is larger than the expected stand-off radius from the ISM alone \cite[$R_{\rm so} \sim 6.7$ kpc,][]{setton_large_2023}, and is most prominent in the Southern hemisphere. MW CGM gas is accelerated to the systemic speed of the LMC in both radial and tangential velocities in the immediate vicinity of the LMC. 
    \item The stripped LMC CGM material is extended along the past orbit of the LMC and is defined by its distinct kinematic signature in Galactocentric radial and tangential velocities in the Southern Hemisphere. Gas stripped earlier from the LMC is moving at radial velocities of $\lesssim -200$ km/s about $100$ kpc from the MW, while more recently stripped gas has large tangential velocities comparable to the systematic speed of the LMC $v_{\rm tan} \gtrsim 300$ km/s. This stripped LMC CGM material is a likely contributor to the ionized budget of the MC Stream, in agreement with the work of \cite{lucchini_magellanic_2020}.
    \item The reflex motion of the MW inner halo with respect to the outer halo produces a global dipole-like response in radial velocities for gas in both hemispheres. Gas $\gtrsim$ 50 kpc is redshifted to radial velocities of $\sim 30-50$ km/s in the Northern hemisphere and blueshifted by a comparable amount in the Southern hemisphere due to the acceleration of the MW towards the past pericentre position of the LMC. This dipole in radial velocities is consistent with other studies that report detections of the reflex motion in the stellar halo. 
    \item Using Mollweide projections of the MW CGM in Galactic coordinates, we find that the CGM response to the infall of the LMC may manifest in key all-sky MW CGM observables of the column density, X-ray surface brightness, and potentially a spatially-varying tSZ Compton-y signal from pressure discontinuities in the MW hot halo. This comes in the form of an enhancement in the column density and X-ray emissivity from stripped LMC CGM gas along the past orbit of the LMC and an additional excess from the shock slightly offset from the LMC, also visible in tSZ across large angular scales in the Southern hemisphere. 
    \item The overall MW CGM response is sensitive to the assumed mass ratio of the MW and LMC CGM. We explore High and Low Mass models of the MW/LMC CGM, ranging from $M_{\rm CGM, MW} = 1.2-6.7 \times 10^{10} M_{\odot}$ and $M_{\rm CGM, LMC} = 1-5.7 \times 10^{9} M_{\odot}$. Lowering the mass ratio by either reducing the initial LMC CGM or increasing the MW CGM mass has the effect of reducing the amount of LMC CGM that survives ram-pressure stripping, which in turn drives a weaker shock (lower Mach number), reduces the mass leftover in the trailing stream, and decreases both the stand-off radius of the shock and the LMC CGM stripping radius. Increasing the mass ratio with either a more massive LMC CGM or a lower mass MW CGM produces the opposite effect. Assuming all of the ionized gas associated with the MC Stream originated in the LMC CGM, then the High Mass LMC CGM leaves a total mass consistent with observations of the ionized gas at an assumed distance of 100 kpc \citep{fox_cosuves_2014}. 
    \item The infall of the LMC has broad implications for the entire MW system. The global dynamical response of the MW CGM is similar to that of the underlying dark matter halo. The LMC shock could affect the morphology and distribution of gaseous structures in the inner MW CGM/ISM. Satellite galaxies of the MW and the LMC that pass through the shock or other overdense regions may also be affected, as the sharp rise in density by factor of 2 across the shock could change the ram-pressure field of the ambient medium and in turn the gas and star formation properties of these systems.  
    \item Most of the LMC CGM gas should be stripped from the ram-pressure of the MW CGM, baring the LMC gas disk to the MW CGM and enabling ram pressure stripping of the ISM \citep[e.g.][]{salem_ram_2015}. This stripped LMC CGM gas is confined almost exclusively to the trailing side of the LMC or beyond $R_{\rm 200, MW}$. This remains the case for all initial MW and LMC CGM masses explored in this work. 
    \item The shock from the LMC may also enhance the synchotron emission of the MW CGM over large angular scales across the Southern hemisphere from the acceleration of cosmic ray electrons and protons. Through diffusive shock acceleration, the LMC shock is predicted to inject a population of cosmic rays with a steep spectral index of $n_E \propto E^{-4}$, producing a small population of synchotron-emitting electrons, however this signal could be enhanced with a pre-shock existing fossil population of cosmic rays. 
\end{itemize}

Satellite-CGM interactions are a common phenomena throughout our universe and knowing how CGM respond to satellite infall, and in turn how that alters its thermodynamics, is a important piece to building new models of galaxies and their atmospheres. The wealth of observations both old and new probing different aspects of the Clouds and their interaction with the Milky Way compels us to ask what signatures from a galaxy’s interaction history are retained in the current structures of the CGM and how these interactions sculpt its physical and kinetic properties. 

\begin{acknowledgments}
CC acknowledges the insightful conversations with J. Colin Hill, David Schiminovich, and Frits Paerels that enhanced this work. GLB acknowledges support from the NSF (AST-2108470 and AST-2307419, ACCESS), a NASA TCAN award, and the Simons Foundation through the Learning the Universe Collaboration. KVJ was supported by Simons Foundation grant 1018465. GB is supported by NSF CAREER AST-1941096.
\end{acknowledgments}

%

\vspace{5mm}







\bibliography{paper}{}
\bibliographystyle{aasjournal}



\end{document}